\documentclass[journal]{IEEEtran}
\usepackage{amsmath}
\usepackage{amsfonts, amssymb}
\usepackage{algorithm}
\usepackage{algorithmic}
\usepackage{graphicx}
\usepackage{epstopdf}
\usepackage{subfigure}
\usepackage{booktabs}
\usepackage{threeparttable}
\usepackage{stfloats}
\usepackage{multirow}
\usepackage[noadjust]{cite}
\usepackage{tabularx}
\usepackage{color}
\usepackage{amsmath} 
\usepackage{amssymb}  
\usepackage{hyperref}

\newtheorem{myPro}{Proposition}
\newtheorem{pf}{Proof}
\newtheorem{myTheo}{Theorem}

\begin{document}

\title{Diffusion Adaptation Framework for Compressive Sensing Reconstruction\thanks{Yicong He, Fei Wang and Badong Chen are with the Institute of Artificial Intelligence and Robotics, Xi'an Jiaotong University, China, e-mails: heyicong@stu.xjtu.edu.cn, wfx@xjtu.edu.cn, chenbd@xjtu.edu.cn.} \thanks{ Shiyuan Wang is with the School of Electronic and Information Engineering, Southwest University, China, e-mail: wsy@swu.edu.cn}}
%
\author{Yicong He, Fei Wang, \IEEEmembership{Member,~IEEE}, Shiyuan Wang, \IEEEmembership{Member,~IEEE}, Badong Chen, \IEEEmembership{Senior~Member,~IEEE}}
%
%
%
%
\maketitle
\begin{abstract}
Compressive sensing(CS) has drawn much attention in recent years due to its low sampling rate as well as high recovery accuracy. As an important procedure, reconstructing a sparse signal from few measurement data has been intensively studied. Many reconstruction algorithms have been proposed and shown good reconstruction performance. However, when dealing with large-scale sparse signal reconstruction problem, the storage requirement will be high, and many algorithms also suffer from high computational cost.
In this paper, we propose a novel diffusion adaptation framework for CS reconstruction, where the reconstruction is performed in a distributed network. The data of measurement matrix are partitioned into small parts and are stored in each node, which assigns the storage load in a decentralized manner. The local information interaction provides the reconstruction ability. Then, a simple and efficient gradient-descend based diffusion algorithm has been proposed to collaboratively recover the sparse signal over network. The convergence of the proposed algorithm is analyzed. To further increase the convergence speed, a mini-batch based diffusion algorithm is also proposed. Simulation results show that the proposed algorithms can achieve good reconstruction accuracy as well as fast convergence speed.
\end{abstract}
%
%
\section{Introduction}
\label{sec:intro}

Compressive sensing(CS) is a novel sampling theory, which provides a new signal sampling (encoding) and reconstruction (decoding) approach \cite{Donoho2006Compressed,candes2006compressive,Candes2006Robust,Candes2004Near,candes2008an}. In detail, given a compressible signal $\boldsymbol{s}=\boldsymbol{\Omega x}$ where $\boldsymbol{\Omega}$ is the transform basis and $\boldsymbol{x}$ is a sparse signal, signal $\boldsymbol{s}$ can be measured by a nonadaptive linear projections (namely sensing matrix) $\boldsymbol{\Phi}$, i.e. $\tilde {\boldsymbol{y}}=\boldsymbol{\Phi s}=\boldsymbol{\Theta x}$ where $\tilde {\boldsymbol{y}}$ is the measurement vector and $\boldsymbol{\Theta}=\boldsymbol{\Phi\Omega}$ is the measurement matrix. Then, at the decoder, $\boldsymbol{x}$ (or $\boldsymbol{s}$) can be recovered from $\tilde {\boldsymbol{y}}$ using reconstruction algorithms. Since CS framework can provide far less sampling rate than Nyquist as well as high recovery accuracy, it has been widely used in many applications such as medical imaging \cite{Lustig2007Sparse} and radar imaging \cite{Potter2010Sparsity}.

As an important procedure in CS, recovering a sparse signal from insufficient number of measurement data has drawn much attention in recent years. In the last decade, many algorithms have been proposed to show accurate reconstruction performance \cite{Mallat1993Matching,Tropp2007Signal,Blumensath2008Iterative,Chen2001Atomic,Chartrand2008Iteratively,Tibshirani2011Regression}. 
An important theoretical guarantee that behind CS reconstruction is the restricted isometry property (RIP) \cite{Cand2008The}. It has been proved that if $\boldsymbol{\Theta}$ obeys RIP, the sparse signal can be recovered from small number of measurements $\boldsymbol{y}$. It has also been shown that random matrices such as Gaussian matrix and Bernoulli matrix can satisfy the RIP condition with a high probability \cite{Cand2006Sparsity,Baraniuk2008A}.

\par Although CS Reconstruction problem has been intensively studied, applying CS reconstruction to large-scale data (such as image data) is still a challenging work. In \cite{Hsieh2015Fast}, the authors proposed a conjugate gradient orthogonal matching pursuit (CG-OMP) algorithm. CG-OMP utilizes Structurally Random Matrix (SRM) \cite{Do2012Fast} as the sensing matrix $\boldsymbol{\Phi}$, which can speed up the signal recovery process as well as reduce the storage requirement. In particular, SRM is related to operator-based approaches, and can improve all greedy algorithms and several iterative shrinkage/threshold (IST) methods such as gradient projection for sparse reconstruction algorithm (GPSR) \cite{Figueiredo2007Gradient} and sparse reconstruction by separable approximation (SpaRSA) \cite{Wright2009Sparse}. However, although $\boldsymbol{\Phi}$ can be fast computed, the transform basis $\boldsymbol{\Omega}$ may not always be fast computable. Thus $\boldsymbol{\Omega}$ or $\boldsymbol{\Theta}$ may still need to be stored, which needs high requirement of storage for large-scale data. To reduce the storage of $\boldsymbol{\Theta}$, a block based compressive sensing (BCS) method was proposed \cite{Gan2007Block,Mun2010Block}. In BCS, the input signal is separated into several small block signals, each signal is individually sensed and recovered. BCS essentially utilizes a block diagonal matrix as the sensing matrix, which, however, is lack of theoretical guarantee. Moreover, BCS needs to modify the sampling strategy at the sensing procedure, and the applicability is limited due to unclear structure.

{\color{black}An efficient way to solve the storage problem for large scale data is to store the data in a distributed network, and then optimize the problem in a distributed way. So far, several distributed optimization algorithms have been proposed including distributed stochastic dual coordinate ascent (DisDCA) \cite{Yang2013Trading}, communication-efficient distributed dual coordinate ascent (CoCoA) \cite{Jaggi2014Communication} and distributed stochastic variance reduced gradient (DSVRG) \cite{Lee2015Distributed}. These algorithms focus on solving the optimization problem in parallel, and need to collect the updated information from all nodes at each iteration. However, in many distributed network such as wireless sensor network (WSN), it is hard to gather the whole information across the network at each iteration.}

\par In this paper we propose a novel diffusion adaptation framework for CS reconstruction. The measurement matrix $\boldsymbol{\Theta}$ are partitioned and stored in a decentralized manner, i.e. each node of the network stores only a small part of $\boldsymbol{\Theta}$. Therefore, the whole storage load is distributed to each node in the network. Further, inspired by diffusion adaptation strategies \cite{Lopes2008Diffusion,Cattivelli2010Diffusion,Chen2012Diffusion,sayed2013diffusion,Abdolee2015Diffusion}, a simple yet efficient diffusion $l_0$-LMS algorithm (D$l_0$-LMS) is proposed. Each node utilizes the finite number of data recursively. The estimation are shared among neighbours at each iteration. Therefore, taking advantages of diffusion adaptation, the D$l_0$-LMS algorithm can collaboratively recover the sparse signals across the network. Utilizing $l_0$-norm as the regularization term can guarantee the sparse estimation. Information exchange within network also provide ability of fast convergence speed, thus greatly increases the computational efficiency. Moreover, a mini-batch based D$l_0$-LMS (MB-D$l_0$-LMS) is also proposed in this study. MB-D$l_0$-LMS utilizes the mini-batch gradient descend (MBGD) method and can further improve the convergence speed.

\par The proposed D$l_0$-LMS is a variant of traditional sparse diffusion LMS algorithm \cite{Chen2009Sparse,Lorenzo2013Sparse}, which have shown abilities in learning the sparse structure over distributed networks. {\color{black}Actually, the formulation of the D-$l_0$LMS algorithm for CS is similar to traditional sparse diffusion algorithm except the data is recursively used. However, applying sparse diffusion algorithm to specific CS task brings more features on both experimental results and theoretical analysis. In particular, diffusion adaptation framework with sparsity constrain gives ability to collaboratively estimate the sparse signal, even though each node cannot individually reconstruct the signal due to insufficient numbers of data. Moreover, in CS all data has been already collected before process, thus the finite known data offers convenience for theoretical analysis, and also gives access to use mini-batch method in gradient estimation. Besides, since the reconstruction speed is a critical issue in evaluating the CS reconstruction algorithm, a larger step size is preferred to achieve faster convergence speed. Obviously the small step size assumption in analysis of traditional diffusion algorithm \cite{Lopes2008Diffusion,Cattivelli2010Diffusion,Chen2012Diffusion,sayed2013diffusion,Abdolee2015Diffusion} is not suitable for analysis for CS, thus in this paper carry out a new theoretical analysis on the step size condition for convergence without small step size assumption.}

\par The proposed D$l_0$-LMS is also related to $l_0$-LMS for CS \cite{Jin2010A}. $l_0$-LMS can be seen as a special case of D$l_0$-LMS where the network contains only 1 node. By introducing traditional sparse LMS algorithm to CS, $l_0$-LMS algorithm has shown great performance improvement compared with other algorithms. In particular, $l_0$-LMS demands less requirement in memory, and achieves better reconstruction performance than other existing algorithms when dealing with large-scale CS reconstruction problem. In D$l_0$-LMS, each node actually performs the same weight update process with $l_0$-LMS, thus the computational complexity of each node in each iteration are the same as $l_0$-LMS. Moreover, the diffusion algorithm gives ability to allow much larger step size for convergence condition, and is confirmed by experiments that the convergence speed is much faster than $l_0$-LMS. Futher, simulations also show that D$l_0$-LMS can achieve similar reconstruction accuracy with $l_0$-LMS.

\par One should also distinguish our work from distributed compressive sensing (DCS) \cite{Baron2005Distributed,Bajwa2006Compressive,Wang2007Distributed,Chen2011Distributed}. In DCS, a number of measurement data are recovered by a group of sensors. The measurement data at each node are assumed to be individually sparse in some basis and are correlated from sensor to sensor. The DCS aims to solve the jointly sparse ensemble reconstruction problem, which is not the topic of our work. Another related work is the Distributed Compressed Estimation(DCE) scheme \cite{Xu2015Distributed}. The DCE incorporates compression and decompression modules into the distributed estimation procedure. The compressed estimator is estimated across the network using diffusion adaptation strategy, and then the reconstruction algorithms are employed to recover the sparse signal from compressed estimator. In DCE, each node still need to store the whole sensing matrix. Moreover, the reconstruction procedure is independent of diffusion adaptation procedure, which may still suffer from the same problem of typical reconstruction methods when dealing with large scale CS reconstruction problem.

\par The paper is organized as follows. In section II we briefly review the concept of compressive sensing and propose the diffusion adaptation framework for CS reconstruction. The gradient based and the mini batch based algorithms for diffusion CS reconstruction are then proposed in Section III. The stability analysis of D$l_0$-LMS is carried out in Section IV. In Section V, simulation results are presented to verify the reconstruction performance. Finally, the conclusion is given in Section VI.

\section{Diffusion Adaptation Framework for Compressive Sensing Reconstruction}
\label{sec:format}

Suppose a real valued discrete signal $\boldsymbol{s} \in \mathbb{R}^{N \times 1}$ is compressible, i.e. $\boldsymbol{s}$ can be represented as $\boldsymbol{s}=\boldsymbol{\Omega}\boldsymbol{x}$ where $\boldsymbol{\Omega}\in \mathbb{R}^{N \times N}$ is a transform basis matrix and $\boldsymbol{x}$ is a sparse signal with sparsity $K\ll N$. In the theory of CS, the sparse signal $\boldsymbol{x}$ can be measured by
\begin{equation}
\tilde {\boldsymbol{y}}=\boldsymbol{\Phi}\boldsymbol{\Omega}\boldsymbol{s}=\boldsymbol{\Theta}\boldsymbol{x}
\end{equation}
where $\boldsymbol{\Theta}=\boldsymbol{\Phi}\boldsymbol{\Omega}$ is the measurement matrix, $\boldsymbol{\Phi}\in \mathbb{R}^{M\times N}$ is the sensing matrix and $\tilde {\boldsymbol{y}}\in \mathbb{R}^{M\times 1}$ is the measurement vector ($M<N$). In practice, the observed measurement vector ${\boldsymbol{y}}$ may be noisy, thus the observed measurement vector can be described as
\begin{equation}
\label{CS1}
\boldsymbol{y}=\boldsymbol{\Theta} \boldsymbol{x}+\boldsymbol{v}
\end{equation}
where $\boldsymbol{v}\in \mathbb{R}^{M\times1}$ is the additive noise vector.
The CS reconstruction task is to recover the sparse signal $\boldsymbol{x}$ from the measurement matrix $\boldsymbol{\Theta}$ and the corresponding noisy measurement $\boldsymbol{y}$. To successfully recover $\boldsymbol{x}$, the measurement matrix $\boldsymbol{\Theta}$ should obey the restricted isometry property (RIP).
\par In practice, the CS reconstruction problem can be viewed as solving a sparse constrained least squares problem with the cost function
\begin{equation}
\label{l0LMS}
J(\boldsymbol{x})= \|\boldsymbol{y}-\boldsymbol{\Theta}\boldsymbol{x}\|_2+\xi\|\boldsymbol{x}\|_0
\end{equation}
where $\|\boldsymbol{x}\|_0$ is the $l_0$ regularization term and $\xi$ is regularization parameter.
\par To apply CS reconstruction in a decentralized manner, one can modify the above cost function. In particular, considering a connected network with $P$ nodes (i.e. the network size is $P$), we can obtain the estimation of $\boldsymbol{x}$ by minimizing the following global cost function
\begin{equation}
\label{Dl0LMS}
{J^{glob}}(\boldsymbol{x}) = \sum_{k=1}^P\|\boldsymbol{y_k}-\boldsymbol{\Theta}_k\boldsymbol{x}\|_2+\xi\|\boldsymbol{x}\|_0
\end{equation}
where
$${\boldsymbol{\Theta }} = \left[ {\begin{array}{*{20}{c}}
{{{\boldsymbol{\Theta }}_1}}\\
{{{\boldsymbol{\Theta }}_2}}\\
 \vdots \\
{{{\boldsymbol{\Theta }}_P}}
\end{array}} \right],{\boldsymbol{y}} = \left[ {\begin{array}{*{20}{c}}
{{{\boldsymbol{y}}_1}}\\
{{{\boldsymbol{y}}_2}}\\
 \vdots \\
{{{\boldsymbol{y}}_P}}
\end{array}} \right]$$
with ${{\boldsymbol{\Theta }}_k}\in{\mathbb{R}}^{L_k\times N}$, ${{\boldsymbol{y}}_k}\in{\mathbb{R}}^{L_k\times 1}$ and $\sum_{k=1}^{P}{L_k}=M$. Since Eq.(\ref{l0LMS}) and Eq.(\ref{Dl0LMS}) are essentially the same cost function, the optimum solution will coincide.
\par
\begin{figure}[tb]
\centering
\includegraphics[width=0.94\linewidth]{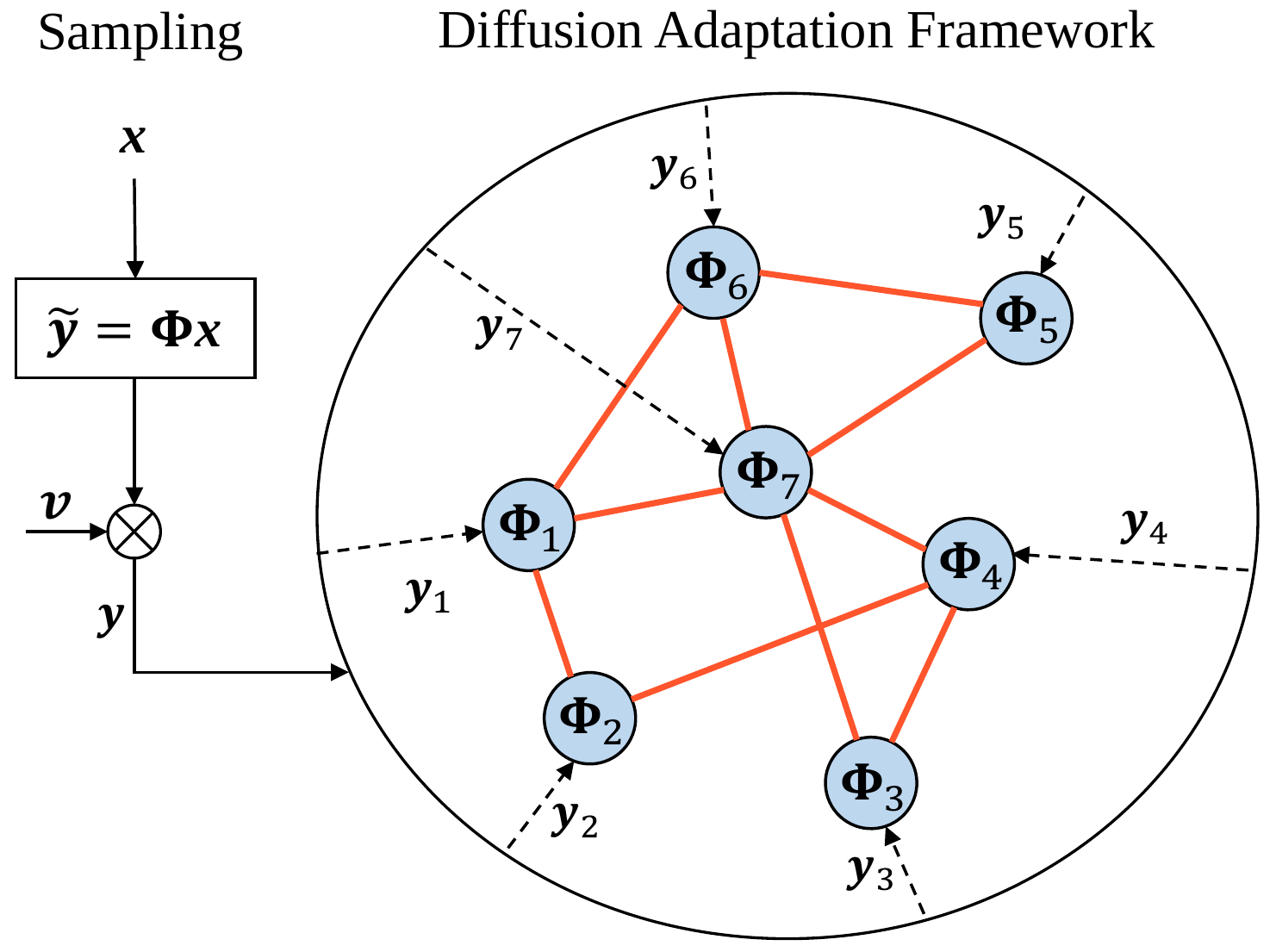}
\caption{Illustration of diffusion adaptation framework for CS reconstruction on a connected network with 7 nodes}
\label{fig1}
\end{figure}

  Fig.\ref{fig1} shows an example of diffusion adaptation framework for CS reconstruction. The connected network includes 7 nodes. The whole ${{\boldsymbol{\Theta}}}$ is partitioned into small parts $\{{\boldsymbol{\Theta}}_k\}_{k=1}^7$. Then, node $k$ only stores a small part of the measurement matrix ${\boldsymbol{\Theta}}_k$ and receives the corresponding measurement data ${{\boldsymbol{y}}_k}$. The information of a node can be transmitted within its neighbourhood (denoted as red links). Although each node has insufficient numbers of measurements and can only exchange information within local neighbours, the information diffusion across the whole network provides the ability to access the whole information of ${{\boldsymbol{\Theta}}}$. 

\section{Proposed Algorithms For Compressive sensing}
\subsection{Gradient descent D$l_0$-LMS for CS}
\label{sec:pagestyle}
\par The diffusion adaptation algorithms for stream data has been intensively studied \cite{Chen2012Diffusion,Lorenzo2013Sparse,Ma2016Diffusion,MaGeneral}. Given the temporal sparse input data sequence $\{\boldsymbol{u}_k(i)\}$ and the corresponding output data sequence $\{\boldsymbol{d}_k(i)\}$ at node $k$, the sparse diffusion LMS adaptation algorithm \cite{Lorenzo2013Sparse} obtains the estimation by minimizing the following global cost function \cite{Lorenzo2013Sparse}
\begin{equation}
\label{DSLMS}
{J^{glob}}(\boldsymbol{w}) = \sum_{k=1}^PE\left[\left({d}_k(i)-\boldsymbol{u}^T_k(i)\boldsymbol{w}\right)^2\right]+\xi\|\boldsymbol{w}\|_0
\end{equation}
Intuitively, we can define $\{\boldsymbol{u}_k(i)\}$ and $\{{d}_k(i)\}$ as
$$\{\boldsymbol{u}_k(i)\}=\{\boldsymbol{\theta}_k(1),\boldsymbol{\theta}_k(2),...,\boldsymbol{\theta}_k(L_k)\}$$
$$\{{d}_k(i)\}=\{y_k(1),y_k(2),...,y_k(L_k)\}$$
where $\boldsymbol{\theta}_k(i)$ and ${y}_k(i)$ denote the transpose of the $i$th row of $\boldsymbol{\Theta}_k$ and the $i$-th scalar of $\boldsymbol{y}_k$, respectively.
Thus the solution to CS reconstruction problem in Eq.({\ref{Dl0LMS}}) can be formulated based on the the diffusion adaptation algorithm {\color{black}with cost function in} Eq.({\ref{DSLMS}}).
\par In traditional diffusion algorithm, the data size $L_k$ is always much larger than input dimension $N$. However, in CS $L_k$ is much smaller than $N$. When directly apply the sparse diffusion algorithm to CS, the adaptation process may not converge to the steady state due to insufficient number of data. To solve this problem, we follow the method described in \cite{Jin2010A} and use the data recursively. In particular, the data used at $i$-th iteration in node $k$ are
\begin{equation}
\begin{aligned}
\label{newsq}
\boldsymbol{u}_k(i)=\boldsymbol{\theta}_k({\rm{mod}}(i,L_k)+1)\\
{d}_k(i)={y}_k({\rm{mod}}(i,L_k)+1)
\end{aligned}
\end{equation}

Therefore, combining diffusion adaptation strategy and modified data sequence in Eq.({\ref{newsq}}), we can derive two gradient-descend based diffusion adaptive algorithms for CS, namely, the Adapt-then-Combine (ATC) diffusion $l_0$-LMS (ATC-D$l_0$-LMS) algorithm
\begin{equation}
\label{ATC}
\renewcommand\arraystretch{1.5}
\left\{ {\begin{array}{*{20}{l}}
{{\boldsymbol{\psi} _k}\left( i+1 \right) = {{\boldsymbol{w}} _k}\left( {i} \right)+{\mu _k}\!\sum\limits_{l \in {{\cal N}_k}}\! {{\alpha _{l,k}}{{\hat {\boldsymbol{g}}}_{l}}\left( {{{\boldsymbol{w}} _k}\left( {i} \right)} \right)}  }\\
~~~~~~~~~~~~~~~~~~~~~-\mu_k\xi\nabla\|\boldsymbol{w}_k(i)\|_0\\
{{{\boldsymbol{w}} _k}\left( {i + 1} \right) = \sum\limits_{l \in {{\cal N}_k}} {{\beta_{l,k}}{\boldsymbol{\psi}_l}\left( {i + 1} \right)} }
\end{array}} \right.
\end{equation}
and the Combine-then-Adapt (CTA) diffusion $l_0$-LMS (CTA-D$l_0$-LMS) algorithm
\begin{equation}
\label{CTA}
\renewcommand\arraystretch{1.5}
\left\{ {\begin{array}{*{20}{l}}
{{\boldsymbol{\varphi} _k}\left( {i} \right) = \sum\limits_{l \in {{\cal N}_k}} {{\beta _{l,k}}{\boldsymbol{w}_l}\left( {i} \right)} }\\
{{\boldsymbol{w}_k}\left( i+1 \right) = {\boldsymbol{\varphi}_k}\left( {i} \right)+{\mu _k}\!\sum\limits_{l \in {{\cal N}_k}}\! {{\alpha _{l,k}}{{\hat {\boldsymbol{g}}}_{l}}\left( {{\boldsymbol{\varphi} _k}\left( {i} \right)} \right)} }\\
~~~~~~~~~~~~~~~~~~~~~-\mu_k\xi\nabla\|\boldsymbol{\varphi}_k(i)\|_0
\end{array}} \right.
\end{equation}
where ${\boldsymbol{\varphi}_k(i)}$ and ${\boldsymbol{\psi} _k(i)}$ are intermediate vectors of node $k$ at time $i$, $\nabla\|\boldsymbol{w}_k(i)\|_0$ is the derivation of $\|\boldsymbol{w}_k(i)\|_0$ and
\begin{equation}
\begin{aligned}
{{{\hat {\boldsymbol{g}}}_{l}}\left( {{{\boldsymbol{x}} _k(i)}} \right)}=\left({d_l(i) - {{\boldsymbol{x}}^T_k(i)}}{{\boldsymbol{u}_l(i)}}\right){{\boldsymbol{u}_l(i)}}
\end{aligned}
\end{equation}
is the instantaneous gradient vector. ${{\cal N}_k}$ is the neighbourhood of node $k$ and $\mu_k$ is the corresponding step size. $\alpha_{l,k},~\beta_{l,k}$ are non-negative weights assigned to link between $l$ and $k$ for adaptation and combination step, respectively. Further, $\alpha_{l,k}, \beta_{l,k}$ can be seen as the $\{l,k\}$-th entries of matrices $\boldsymbol{S}$ and $\boldsymbol{A}$ respectively. Specifically, $\boldsymbol{S}$ and $\boldsymbol{A}$ should satisfy
\begin{equation}
\label{A1}
\boldsymbol{S}\boldsymbol{1}=\boldsymbol{A}^T\boldsymbol{1}=\boldsymbol{1}
\end{equation}
and
\begin{equation}
\label{A2}
\boldsymbol{S}(l,k)=\boldsymbol{A}(l,k)=0~~if~~ l\notin {\cal N}_k
\end{equation}
where $\boldsymbol{1}$ denotes the all one vector.

\par An important problem left is to calculate $\nabla\|\boldsymbol{w}(i)\|_0$. Since $\|\boldsymbol{w}(i)\|_0$ is non-differentiable, one should use an approximation instead. There are several approximations of $l_0$ norm \cite{Weston2003Use} which can work well for sparse identification \cite{Jin2010A,Lorenzo2013Sparse}. In this paper we use the zero attraction term $\boldsymbol{z}_{\delta}(\boldsymbol{w}_k(i))$ similar to $l_0$-LMS, which is defined as \cite{Jin2010A}
\begin{equation}
\begin{aligned}
-\nabla\|\boldsymbol{w}(i)\|_0&\approx \boldsymbol{z}_{\delta}(\boldsymbol{w}(i))\\
&=[z_{\delta}(w_1(i)),z_{\delta}(w_2(i)),...,z_{\delta}(w_N(i))]^T
\end{aligned}
\end{equation}
where
\begin{equation}
\label{ZA}
\renewcommand\arraystretch{1.2}
z_{\delta}(w_m(i))=\left\{ {\begin{array}{*{20}{c}}
\delta^2w_m(i)+\delta~~ -1/\delta\leq w_m(i)<0\\
\delta^2w_m(i)-\delta~~~~~ 0<w_m(i)\leq 1/\delta\\
~~~~~~0~~~~~~~~~~~~~~ |w_k(i)|>1/\delta
\end{array}} \right.
\end{equation}
and $\delta$ is the zero attraction parameter.
\par The pesudo code of ATC-D$l_0$-LMS is summarized in Algorithm 1. At iteration $i$, each node $k$ sends the data pair $\{\boldsymbol{u}_k(i),d_k(i)\}$ to its neighbours. Then the adaptation is performed at each node. After adaptation, the estimation in each node is transferred to its neighbours for combination. The process of CTA-D$l_0$-LMS is similar with ATC-D$l_0$-LMS except that the order of adaptation step and combination step are reversed.\\

\begin{algorithm}
\caption{ATC-D$l_0$-LMS Algorithm for CS}
\begin{algorithmic}
 \STATE \emph{Initialization}
 \STATE Choose step-size $\mu_k$ for each node $k$, regularization parameter $\xi$, zero attraction parameter $\delta$ and maximum iteration number $C$. Set initial iteration number $i=1$ and weight vector $\boldsymbol{w}_k(1)=\boldsymbol{0}$ for all node $k$. Select $\alpha_{l,k},~\beta_{l,k}$ according to Eq.(\ref{A1}) and Eq.(\ref{A2}).
 {\color{black}\STATE \emph{Data Partition}
 \STATE Assign the data of measurement matrix $\boldsymbol{\Theta}$ and corresponding $\boldsymbol{y}$ to each node according to partition strategy.}
 \STATE \emph{Computation}
 \WHILE {$i<C$}
 \FOR {each node $k$}
 \STATE $n={\rm{mod}}(i,L_k)+1,\boldsymbol{u}_k(i)=\boldsymbol{\theta}_k(n),d_k(i)=y_k(n)$
 \STATE \textbf{Communication 1:}
 \STATE Transmit $\{\boldsymbol{u}_k(i),d_k(i)\}$ to neighbour node $l$ in ${\cal N}_k$
 \STATE \textbf{Adaptation:}
 \STATE ${{\boldsymbol{\psi} _k}\!\left( i+1 \right) = {{\boldsymbol{w}} _k}\!\left( {i} \right)+{\mu _k}\!\sum\limits_{l \in {{\cal N}_k}}\! {{\alpha _{l,k}}{{\hat {\boldsymbol{g}}}_{l}}\!\left( {{{\boldsymbol{w}} _k}\left( {i} \right)} \right)}  }$\\$~~~~~~~~~~~~~~~~~~~~~+\mu_k\xi\boldsymbol{z}_{\delta}(\boldsymbol{w}(i))$
 \STATE \textbf{Communication 2:}
 \STATE Transmit ${{\boldsymbol{\psi} _k}\!\left( i+1 \right)}$ to neighbour node $l$ in ${\cal N}_k$
 \STATE \textbf{Combination:}
 \STATE ${{{\boldsymbol{w}} _k}\left( {i + 1} \right) = \sum\limits_{l \in {{\cal N}_k}} {{\beta_{l,k}}{\boldsymbol{\psi}_l}\left( {i + 1} \right)} }$
 \ENDFOR
 \IF {the stop criterion is satisfied}
 \STATE break
 \ENDIF
 \STATE \%update iteration number
 \STATE $i=i+1$
\ENDWHILE
\STATE \emph{Output:} ${\boldsymbol{w}}=\frac{1}{P}\sum_{k=1}^P{{\boldsymbol{w}} _k}(i)$
\end{algorithmic}
\end{algorithm}

{\bf{Remark 1}}: Consider a special case that the network size $P=1$. In this case, the reconstruction is processed in a stand-alone manner, and Eq.({\ref{ATC}}) and Eq.({\ref{CTA}}) will be equal to
\begin{equation}
{{\boldsymbol{w}}\!\left( i+1 \right) \!=\! {{\boldsymbol{w}}}\left( {i} \right)\!+\!{\mu}\! \left({d(i) \!-\! {{\boldsymbol{x}}^T\!(i)}}{{\boldsymbol{u}(i)}}\right)\!{{\boldsymbol{u}(i)}} \!+\!\mu\xi\boldsymbol{z}_{\delta}(\boldsymbol{w}(i))}\\
\end{equation}
which is the typical $l_0$-LMS for CS \cite{Jin2010A}. Therefore, $l_0$-LMS can be viewed as a special case of ATC-D$l_0$-LMS and CTA-D$l_0$-LMS.

{\color{black}{\bf{Remark 2}}: In a strict sense, the name D$l_0$-LMS is nonstandard since the algorithm actually minimize an approximation of $l_0$ norm. In this paper, we just use the D$l_0$-LMS to keep the name consistent with $l_0$-LMS for CS \cite{Jin2010A}.}

{\bf{Remark 3}}: For large scale data, to reduce the amount of network transmission, one can put away the adaptation information exchange, i.e. $\boldsymbol{S}=\boldsymbol{I}$ \cite{fernandez2017adaptive}. That is, each node only utilize its own data to perform adaptation, and then share its estimation to neighbours for combination. Simulation results in Section V will verify the feasibility of this strategy.

\subsection{Mini-batch D$l_0$-LMS for CS}
The proposed gradient-based D$l_0$-LMS is a typical extension of traditional sparse diffusion algorithm. We should notice that unlike traditional diffusion adaptation, in CS all the data is already known. Therefore, one can utilize more information during each iteration. In \cite{Jin2010A}, the $l_0$ regularized exponetially forgetting windows LMS ($l_0$-EFWLMS) algorithm is proposed to improve the convergence speed of $l_0$-LMS. Extended from affine projection algorithm(APA) \cite{Member1984An}, $l_0$-EFWLMS utilizes a sliding window approach to use more data to improve the gradient estimation. However, $l_0$-EFWLMS still follows the traditional adaptive filtering method.
\par In data regression problem, mini-batch gradient descent (MBGD) method has been widely used. MBGD selects a small part of the sample data, computes gradient for each data, and then calculates the average gradient as the gradient estimation. For diffusion CS, the input data for node $k$ at each iteration $i$ can be chosen as
\begin{equation}
\begin{aligned}
\label{UD}
\boldsymbol{U}_k(i)=[\boldsymbol{\theta}(r(1)),\boldsymbol{\theta}(r(2)),...,\boldsymbol{\theta}(r(Q))]^T\\
\boldsymbol{D}_k(i)=[y(r(1)),y(r(2)),...,y(r(Q))]^T\\
\end{aligned}
\end{equation}
where $\boldsymbol{r}=[r(1)~r(2)~...~r(Q)]^T\in\mathbb{N}_{+}^{Q\times1},Q\leq\min\{L_k\}$ is the index vector whose elements are uniformly and randomly chosen from $[1,M]$.
Then, according to MBGD method, the average gradient ${{{\hat {\boldsymbol{G}}}_{l}}\left( {{{\boldsymbol{x}} _k(i)}} \right)}$ is defined as
\begin{equation}
\begin{aligned}
\label{Gl}
{{\hat{{\boldsymbol{G}}}_{l}}\left( {{{\boldsymbol{x}} _k(i)}} \right)}=\frac{1}{Q}{{\boldsymbol{U}_l(i)}}\left({\boldsymbol{D}_l(i) - {{\boldsymbol{x}}^T_k(i)}}{{\boldsymbol{U}_l(i)}}\right)
\end{aligned}
\end{equation}
Then, the weight update of corresponding ATC mini-batch diffusion $l_0$-LMS algorithm (ATC-MB-D$l_0$-LMS) can be represented as
\begin{equation}
\label{MBATC}
\renewcommand\arraystretch{1.5}
\left\{ {\begin{array}{*{20}{l}}
{{\boldsymbol{\psi} _k}\left( i+1 \right) = {{\boldsymbol{w}} _k}\left( {i} \right)+{\mu _k}\!\sum\limits_{l \in {{\cal N}_k}}\! {{\alpha _{l,k}}{\hat{{\boldsymbol{G}}}_{l}}\left( {{{\boldsymbol{w}} _k}\left( {i} \right)} \right)}  }\\
~~~~~~~~~~~~~~~~~~~~~+\mu_k\xi\boldsymbol{z}_{\delta}(\boldsymbol{w}(i))\\
{{{\boldsymbol{w}} _k}\left( {i + 1} \right) = \sum\limits_{l \in {{\cal N}_k}} {{\beta_{l,k}}{\boldsymbol{\psi}_l}\left( {i + 1} \right)} }
\end{array}} \right.
\end{equation}
where the instantaneous gradient ${{{\hat {\boldsymbol{g}}}_{l}}\left( {{{\boldsymbol{x}} _k(i)}} \right)}$ in Eq.(\ref{ATC}) is replaced by ${{{\hat {\boldsymbol{G}}}_{l}}\left( {{{\boldsymbol{x}} _k(i)}} \right)}$. The corresponding CTA mini-batch diffusion $l_0$-LMS algorithm (CTA-MB-D$l_0$-LMS) can be also simply derived from Eq.(\ref{CTA}).
\par In real application, utilizing mini-batch method gives faster convergence speed than D$l_0$-LMS, but may also cause instability when the step size is large. To alleviate the negative impact caused by MBGD method, we optimize the iterative process by constraining the sparsity variance during the convergence process. In particular, the sparsity of the estimation in at $i$-th iteration is defined as
\begin{equation}
\label{Stop2}
s(\boldsymbol{w}(i))=\sum_{j=1}^{N} f(w_j(i))
\end{equation}
where
\[
f\left( x \right) = \left\{ {\begin{array}{*{20}{c}}
{1,\left| x \right| > \tau }\\
{0,\left| x \right| < \tau }
\end{array}} \right.
\]
and $\tau$ is a small positive threshold. Then, after a small number of iterations $0.02N$, if the sparsity of $i+1$-th iteration is larger than 1.5 times of the sparsity at $i$-th iteration, the estimation will not update.
\par The pesudo code of ATC-MB-D$l_0$-LMS is summarized in Algorithm 2. Unlike Algorithm 1 where $\{\boldsymbol{u}_k(i),d_k(i)\}$ is transmitted within neighbours, to alleviate the load of network transmission, here the estimation $\boldsymbol{w}_k(i)$ is transmitted to neighbours. The gradient is computed at neighbour nodes and then sent back. Similar to ATC-D$l_0$-LMS, one can also set $\boldsymbol{S}=\boldsymbol{I}$ to put away the adaptation step to save the amount of network transmission.

\begin{algorithm}
\caption{ATC-MB-D$l_0$-LMS Algorithm for CS}
\begin{algorithmic}
 \STATE \emph{Initialization}
 \STATE Choose step-size $\mu$, regularization parameter $\xi$, zero attraction parameter $\delta$, maximum iteration number $C$ , threshold $\tau$ and mini-batch size $Q$. Set initial iteration number $i=1$ and weight vector $\boldsymbol{w}_k(1)=\boldsymbol{0}$ for all node $k$. Select $\alpha_{l,k},~\beta_{l,k}$ according to Eq.(\ref{A1}) and Eq.(\ref{A2}).
 {\color{black}\STATE \emph{Data Partition}
 \STATE Assign the data of measurement matrix $\boldsymbol{\Theta}$ and corresponding $\boldsymbol{y}$ to each node according to partition strategy.}
 \STATE \emph{Computation:}
 \WHILE {$i<C$}
 \FOR {each node $k$}
 \STATE {Select index vector $\boldsymbol{r}$, then generate $\boldsymbol{U}_k(i)$ and $\boldsymbol{D}_k(i)$ from Eq.({\ref{UD}}})
 \STATE \textbf{Communication 1:}
 \STATE Transmit $\boldsymbol{w}_k(i)$ to neighbour node $l$ in ${\cal N}_k$
 \STATE \textbf{Gradient estimation:}
 \STATE Compute ${{{{\boldsymbol{G}}}_{k}}\left( {{{\boldsymbol{w}} _l(i)}} \right)}$ from Eq.({\ref{Gl}}) for all $l\in{\cal N}_k$
 \STATE \textbf{Communication 2:}
 \STATE Transmit ${{{{\boldsymbol{G}}}_{k}}\left( {{{\boldsymbol{w}} _l(i)}} \right)}$ to neighbour node $l$
 \STATE \textbf{Adaptation:}
 \STATE ${{\boldsymbol{\psi} _k}\!\left( i+1 \right) = {{\boldsymbol{w}} _k}\!\left( {i} \right)+{\mu _k}\!\sum\limits_{l \in {{\cal N}_k}}\! {{\alpha _{l,k}}{{ {\boldsymbol{G}}}_{l}}\!\left( {{{\boldsymbol{w}} _k}\left( {i} \right)} \right)}  }$\\$~~~~~~~~~~~~~~~~~~~~~+\mu_k\xi\boldsymbol{z}_{\delta}(\boldsymbol{w}(i))$
 \STATE \textbf{Communication 3:}
 \STATE Transmit ${{\boldsymbol{\psi} _k}\!\left( i+1 \right)}$ to neighbour node $l$ in ${\cal N}_k$
 \STATE \textbf{Combination:}
 \STATE ${{{\boldsymbol{w}} _k}\left( {i + 1} \right) = \sum\limits_{l \in {{\cal N}_k}} {{\beta_{l,k}}{\boldsymbol{\psi}_l}\left( {i + 1} \right)} }$
 \IF {$s(\boldsymbol{w}_k\!(i+1))\!-\!s(\boldsymbol{w}_k(i))\!>\!1.5s(\boldsymbol{w}_k(i))~ {\rm{and}} ~i\!>\!0.02N$}
 \STATE $\boldsymbol{w}_k(i+1)=\boldsymbol{w}_k(i)$
 \ENDIF
 \ENDFOR
 \IF {the stop criterion is satisfied}
 \STATE break
 \ENDIF
 \STATE \%update iteration number
 \STATE $i=i+1$
\ENDWHILE
\STATE \emph{Output:} ${\boldsymbol{w}}=\frac{1}{P}\sum_{k=1}^P{{\boldsymbol{w}} _k}(i)$
\end{algorithmic}
\end{algorithm}

\subsection{Data Partition Strategy and Stop criterion}
{\color{black}For a connected network, using diffusion based algorithm allows each node to observe all the information of the data, thus in theory for any partition strategy the sparse signal can always be recovered. However, as discussed later in Part E, Section IV, the data correlation may influence the convergence condition on step size which is directly related to reconstruction speed. Since the data is recursively used, the data correlation may occur every $L_k$ iterations. Therefor, to avoid high correlation of data, the partition strategy should selected so that $L_k$ for each node $k$ are as large as possible. In practice, uniformly assign the data to each node is a proper choice.}

Although one can use the maximum iteration number $C$ to stop the iteration, one would like a more practical stop criterion. In \cite{Jin2010A}, the author utilizes the squared error between adjacent estimation as the index of stop condition. However, it is always not operational in real applications. Observing the fact that the sparsity of the estimation will maintain as the algorithm converges to the steady state, here we propose a new stop criterion based on the sparsity of the estimation: given the window length $L_s$ and the threshold $p_s$, by defining the count at iteration $i$
\begin{equation}
\label{Stop1}
S_c(i)=\sum_{j=i-L_s+1}^{i} p(s(\boldsymbol{w}(j)))
\end{equation}
where
\[p\left( x \right) = \left\{ {\begin{array}{*{20}{c}}
{1,s_{min} \leq x  \leq s_{min}+p_s}\\
{0,x  > s_{min}+p_s }
\end{array}} \right.\]
and $s_{min}=\min\{s(\boldsymbol{w}(j))\}_{j=i-L_s+1}^{i}$, if $S_c(i)>0.8L_s$, then the algorithm will stop.

\section{Convergence Analysis}
In this section, we carry out the convergence analysis of D$l_0$-LMS algorithm for CS reconstruction. We first derive the recursion form of the algorithm, and then analyze the sufficient condition for convergence on step size under different parameter settings. Finally the influence of regularization term and data reuse is also discussed.

\subsection{Recursion form derivation}

For simplicity, similar to \cite{Chen2012Diffusion,Lorenzo2013Sparse}, we carry out the mean-square analysis on the following general diffusion framework structure
\begin{equation}
\label{generaldiffusion}
\left\{ {\begin{array}{*{20}{c}}
{{\boldsymbol{\varphi} _k}\left( {i} \right) = \sum\limits_{l \in {{\cal N}_k}} {{\beta _{l,k}}{\boldsymbol{w}_l}\left( {i} \right)} }\\
\!\!{{\boldsymbol{\theta} _k}\left( i\!+\!1 \right) \!=\! {\boldsymbol{\varphi} _k}\left( {i} \right)\!+\!{\mu _k}\!\!\sum\limits_{l \in {{\cal N}_k}} {{\alpha _{l,k}}{{\hat {\boldsymbol{g}}}_{l}}\left( {{\boldsymbol{\varphi} _k}\left( {i} \right)} \right)} }+\mu_k\xi\boldsymbol{z}_{\delta}(\boldsymbol{w}(i))\\
{{\boldsymbol{w}_k}\left( i+1 \right) = \sum\limits_{l \in {{\cal N}_k}} {{\gamma _{l,k}}{\boldsymbol{\theta} _l}\left( i+1 \right)} }
\end{array}} \right.
\end{equation}
where $\alpha_{l,k}, \beta_{l,k}, \gamma_{l,k}$ can be seen as the $\{l,k\}$-th entries of matrices $\boldsymbol{A}_1$ and $\boldsymbol{S}$ and $\boldsymbol{A}_2$ respectively. ATC-D$l_0$-LMS and CTA-D$l_0$-LMS can be viewed as two special cases by setting ${\boldsymbol{A}}_1=\boldsymbol{I}$ and ${\boldsymbol{A}}_2=\boldsymbol{I}$, respectively.

Subtracting both sides of Eq.(\ref{generaldiffusion}) from the desired sparse vector $\boldsymbol{x}$, we can obtain
\begin{equation}
\label{DMTCh}
\left\{ {\begin{array}{*{20}{c}}
{{\tilde{\boldsymbol{\varphi}} _k}\left( {i} \right) = \sum\limits_{l \in {{\cal N}_k}} {{\beta _{l,k}}{\tilde{\boldsymbol{w}}_l}\left( {i} \right)} }\\
\!\!{{\tilde{\boldsymbol{\theta}} _k}\left( i\!+\!1 \right) \!=\! {\tilde{\boldsymbol{\varphi}} _k}\left( {i} \right)-\!{\mu _k}\!\!\sum\limits_{l \in {{\cal N}_k}} {{\alpha _{l,k}}{{\hat {\boldsymbol{g}}}_{l}}\left( {{\boldsymbol{\varphi} _k}\left( {i} \right)} \right)} }-\mu_k\xi\boldsymbol{z}_{\delta}(\boldsymbol{w}(i))\\
{{\tilde{\boldsymbol{w}}_k}\left( i+1 \right) = \sum\limits_{l \in {{\cal N}_k}} {{\gamma _{l,k}}{\tilde{\boldsymbol{\theta}} _l}\left( i+1 \right)} }
\end{array}} \right.
\end{equation}
where ${\tilde{\boldsymbol{\varphi}} _k}\left( {i} \right) = \boldsymbol{x} - {{\boldsymbol{\varphi}} _k}(i)$, $\tilde {\boldsymbol{w}}_k(i) = \boldsymbol{x} - \boldsymbol{w}_k(i)$, ${\tilde{\boldsymbol{\theta}} _k}\left( i \right) = \boldsymbol{x} - \boldsymbol{\theta}_k(i)$ are error vectors. Therefore, by defining
\begin{gather*}
{\boldsymbol{w}}(i)={\rm{col}}\{{\boldsymbol{w}}_k(i)\}_{k=1}^{P}, \tilde {\boldsymbol{w}}(i)={\rm{col}}\{\tilde {\boldsymbol{w}}_k(i)\}_{k=1}^{P}\\
\boldsymbol{\cal A}_1=\boldsymbol{A}_1 \otimes \boldsymbol{I}_N, \boldsymbol{\cal A}_2=\boldsymbol{A}_2 \otimes \boldsymbol{I}_N, \boldsymbol{\cal S}=\boldsymbol{S} \otimes \boldsymbol{I}_N\\
\boldsymbol{D}={\rm diag}\{\mu_k\}_{k=1}^{P},{\boldsymbol{\cal D}}=\boldsymbol{D}\otimes\boldsymbol{I}_N\\
\boldsymbol{V}(i)={\rm diag}\{v_k(i)\}_{k=1}^{P},{\boldsymbol{\cal V}(i)}=\boldsymbol{V}(i)\otimes\boldsymbol{I}_N\\
\boldsymbol{\cal H}(i) =\sum\nolimits_{l = 1}^P {\rm diag}\{\alpha_{l,k}{{\boldsymbol{u}_l(i)}}{{\boldsymbol{u}_l(i)^T}}\}_{k=1}^{P}\\
\boldsymbol{s}(i)=\boldsymbol{\cal S}^T\boldsymbol{\cal V}^T(i){\rm{col}}\{{\boldsymbol{u}}_k(i)\}_{k=1}^{P}
\end{gather*}
one can obtain the recursion from Eq.(\ref{DMTCh})
\begin{equation}
\label{EMTC3}
\tilde {\boldsymbol{w}}(i + 1)\! =\! \boldsymbol{\cal A}_2^T[\boldsymbol{I}-\boldsymbol{\cal D}\boldsymbol{\cal H}]\boldsymbol{\cal A}_1^T\tilde {\boldsymbol{w}}(i) - \boldsymbol{\cal A}_2^T\boldsymbol{\cal D}\boldsymbol{s}(i) - \xi {\boldsymbol{\cal A}}_2^T{\boldsymbol{{\cal D}z}_{\delta}}\!\left( {{\boldsymbol{w}}\!\left( i \right)} \right)
\end{equation}

\subsection{Sufficient condition for convergence under deterministic measurement matrix}

{\color{black}Based on the recursion in Eq.({\ref{EMTC3}}) and property that the data is recursively used, one can further derive the following theorem.
\begin{myPro}
Given a certain measurement matrix $\boldsymbol{\Theta}$ and matrices $\boldsymbol{A}_1$, $\boldsymbol{S}$, $\boldsymbol{A}_2$, for any finite initial vector $\boldsymbol{w}(1)$ and finite noise vector $\boldsymbol{v}$, define the product
\begin{equation}
\label{Gamma}
\boldsymbol{\Gamma}=\prod\limits_{n = 1}^{L_{m}} {{\boldsymbol{\cal A}}_1^T\left[ {{\boldsymbol{I}} - {\boldsymbol{\cal DH}}\left( n \right)} \right]{{\boldsymbol{\cal A}}_2}}
\end{equation}
where $L_m$ is the least common multiple of $\{L_k\}_{k=1}^P$, and define the sequence $\{\lambda_i\}_{i=1}^{N\times P}$ as the eigenvalues of $\boldsymbol{\Gamma}$ arranged from large to small according to the modulus. For arbitrary initial condition, in the diffusion algorithm in Eq.(\ref{EMTC3}) will converge if the step sizes $\mu_{k}$ are selected so that the modulus of the $(N-M+1)$-th eigenvalue is less than 1, i.e.
\begin{equation}
\label{T1}
|\lambda_{N-M+1}|<1
\end{equation}
\end{myPro}
\begin{pf}
See Appendix A.
\end{pf}

The above result can be used in practical since the generated measurement matrix $\boldsymbol{\Theta}$ is always fixed during the specific CS task. Different from traditional diffusion process, once the data has been collected, the process is deterministic and there is no more randomness in the operation of the reconstruction.}

\subsection{Sufficient condition for convergence on random measurement matrices}
In many situations, the measurement matrix may not always fixed during all reconstruction process. However, since the measurement matrix is always selected as the random matrix, we can follow the method of traditional diffusion algorithm \cite{Lopes2008Diffusion,Cattivelli2010Diffusion,Chen2012Diffusion,sayed2013diffusion,Abdolee2015Diffusion} and analyze the convergence of the proposed algorithm in mean and mean squares sense.

Similar to traditional diffusion algorithms, for tractability of the analysis we use the following assumptions:\\
\textbf{A1}. The elements of noise vector $\boldsymbol{v}$ are i.i.d processes and independent of measurement matrix $\boldsymbol{\Theta}$.\\
\textbf{A2}. $L_k$ of each node is sufficient large so that $\boldsymbol{w}_k(i)$ at arbitrary node $k$ is independent of $\boldsymbol{u}_l(i),l\in{\cal N}_k$.\\
\textbf{A3}. The noise has finite variance.\\

{\color{black}We should remark that in CS reconstruction, due to recursively use of data, Assumption A2 may not be satisfied in practice. Nevertheless, if $L_k$ is sufficient large so that the correlation is sufficient small, the independent assumption can almost reach. This fact has been proved by simulation results of $l_0$-LMS for CS \cite{Jin2010A}. Moreover, one should also notice that the small step size assumption in analysis of traditional diffusion algorithms is removed. In \cite{Lopes2008Diffusion,Cattivelli2010Diffusion,Chen2012Diffusion,sayed2013diffusion,Abdolee2015Diffusion}, the analysis is based on the assumption that the step sizes are sufficient small so that the higher-order powers of step sizes can be ignored. However, in CS a larger step size is preferred to achieve faster convergence speed. Thus in this paper we eliminate this assumption and propose a new analysis on mean and mean squares performance. In the analysis, without loss of generality, the variances for each element of $\boldsymbol{\Theta}$ are all set to $1/M$.}


Postmultipling both sides of Eq.({\ref{EMTC3}}) with their respective transposes, and then utilizing trace operation, we can obtain the following weighted mean square relation
\begin{equation}
\label{EMTC5}
\begin{aligned}
\!E\!\left[ {\left\| {{\boldsymbol{\tilde w}}\!\left( {i \!+\! 1} \right)} \right\|_{\boldsymbol{\Sigma}} ^2} \right] \!\!= &E\left[ {\left\| {{\boldsymbol{\tilde w}}\left( {i \!+\! 1} \right)} \right\|_{\boldsymbol{\Lambda}} ^2} \right]+ tr\left\{\Sigma {\boldsymbol{U}(i)}\right\}
\end{aligned}
\end{equation}
where
\begin{equation}
\label{EMTC6}
\begin{aligned}
{\boldsymbol{U}(i)}= & {\xi ^2} {\boldsymbol{\cal A}}_2^T{\boldsymbol{{\cal D}K}}\left( i \right){\boldsymbol{\cal D}}{{\boldsymbol{\cal A}}_2}+  {\boldsymbol{\cal A}}_2^T{\boldsymbol{\cal D}}{{\boldsymbol{\cal S}}^T}{\boldsymbol{P}}\left( i \right){\boldsymbol{\cal SD}}{{\boldsymbol{\cal A}}_2}\\
 &-\xi  {\boldsymbol{\cal A}}_2^T\left( {{\boldsymbol{I}} - \frac{1}{M}{\boldsymbol{{\cal DG}}}} \right){\boldsymbol{\cal A}}_1^T{\boldsymbol{J}}\left( i \right){\boldsymbol{\cal D}}{{\boldsymbol{\cal A}}_2}\\
 &- \xi  {\boldsymbol{\cal A}}_2^T{\boldsymbol{{\cal D} J}}\left( i \right){{\boldsymbol{\cal A}}_1}\left( {{\boldsymbol{I}} - \frac{1}{M}{\boldsymbol{\cal DG}}} \right){{\boldsymbol{\cal A}}_2}
\end{aligned}
\end{equation}
\begin{equation}
\begin{aligned}
\boldsymbol{\Lambda}  = & {{\boldsymbol{\cal A}}_1}\left( {{\boldsymbol{I}} - \frac{1}{{{M}}}{\boldsymbol{\cal GD}}} \right){{\boldsymbol{\cal A}}_2}\boldsymbol{\Sigma} {\boldsymbol{\cal A}}_2^T\left( {{\boldsymbol{I}} - \frac{1}{{{M}}}{\boldsymbol{\cal DG}}} \right){\boldsymbol{\cal A}}_1^T \\
&\!+ \!\frac{{N \!+\! 1}}{{{M^2}}}\sum\limits_{k = 1}^P {{\boldsymbol{\cal A}}_1\boldsymbol{\cal T}_k{\boldsymbol{\cal D}}{{\boldsymbol{\cal A}}_2}\boldsymbol{\Sigma} {\boldsymbol{\cal A}}_2^T{{\boldsymbol{\cal D}}}\boldsymbol{\cal T}_k{{\boldsymbol{\cal A}}_1^T}}
\end{aligned}
\end{equation}
with
\begin{equation}
\begin{aligned}
\boldsymbol{G} =\sum\nolimits_{l = 1}^P {\rm diag}&\{\alpha_{l,k}\}_{k=1}^{P},{\boldsymbol{\cal G}}=\boldsymbol{G}\otimes\boldsymbol{I}_N\\
\boldsymbol{T}_k={\rm diag}&\{\alpha_{l,k}\}_{l=1}^{P},\boldsymbol{\cal T}_k = { \boldsymbol{T}_k \otimes {{\boldsymbol{I}}_N}}\\
{\boldsymbol{J}}\left( i \right) &= E\left[ {{\boldsymbol{\tilde w}}\left( i \right){{\boldsymbol{z}}^T}\left( {{\boldsymbol{w}}\left( i \right)} \right)} \right]\\
{\boldsymbol{K}}\left( i \right) &= E\left[ {{\boldsymbol{z}}\left( {{\boldsymbol{w}}\left( i \right)} \right){{\boldsymbol{z}}^T}\left( {{\boldsymbol{w}}\left( i \right)} \right)} \right]\\
{\boldsymbol{P}}\left( i \right) &= E\left[ {{\boldsymbol{\cal V}}{{\left( i \right)}^T}{\boldsymbol{\cal V}}\left( i \right)} \right]
\end{aligned}
\end{equation}
Using the relationship of vectorization operator, matrix trace and Kronecker product
\begin{equation}
\begin{aligned}
vec\{\boldsymbol{ABC}\}&=(\boldsymbol{C}^T\otimes\boldsymbol{A})vec\{\boldsymbol{B}\}\\
tr\{\boldsymbol{A}^T\boldsymbol{B}\}&=vec\{\boldsymbol{B}\}^Tvec\{\boldsymbol{A}\}
\end{aligned}
\end{equation}
by defining $\boldsymbol{\sigma}=vec\{\boldsymbol{\Sigma}\}$, Eq.(\ref{EMTC5}) can be derived as
\begin{equation}
\begin{aligned}
\!E\!\left[ {\left\| {{\boldsymbol{\tilde w}}\!\left( {i \!+\! 1} \right)} \right\|_{\boldsymbol{\sigma}} ^2} \right] \!\!= &E\left[ {\left\| {{\boldsymbol{\tilde w}}\left( {i \!+\! 1} \right)} \right\|_{\boldsymbol{{\cal F}\sigma}} ^2} \right]+{\rm{vec}}\{\boldsymbol{U}^T(i)\}^T\boldsymbol{\sigma}
\end{aligned}
\end{equation}
where ${\left\| {{\boldsymbol{\tilde w}}\!\left( {i \!+\! 1} \right)} \right\|_{\boldsymbol{\sigma}} ^2}$ and ${\left\| {{\boldsymbol{\tilde w}}\!\left( {i \!+\! 1} \right)} \right\|_{\boldsymbol{\Sigma}} ^2}$ denotes the same quantity, and
\begin{equation}
\label{calF}
\begin{aligned}
\boldsymbol{\cal F}=&\left( {{{\boldsymbol{\cal A}}_1} \otimes {{\boldsymbol{\cal A}}_1}} \right)\left[ {\left({\boldsymbol{I}} - \frac{1}{{{M}}}{\boldsymbol{\cal GD}} \right)\otimes \left({\boldsymbol{I}} - \frac{1}{{{M}}}{\boldsymbol{\cal GD}} \right)}\right.\\
&+ \left.{\frac{{N + 1}}{{{M^2}}}\sum\limits_{k = 1}^M {\left( {{{\boldsymbol{\cal T}}_k}{\boldsymbol{\cal D}} \otimes {{\boldsymbol{\cal T}}_k}{\boldsymbol{\cal D}}} \right)} } \right)\left( {{{\boldsymbol{\cal A}}_2} \otimes {{\boldsymbol{\cal A}}_2}} \right)
\end{aligned}
\end{equation}
\par It is easy to prove that $\boldsymbol{K}(i)$ and $\boldsymbol{P}(i)$ are always bounded. Moreover, for $\boldsymbol{J}(i)$ one can obtain
\begin{equation}
|\boldsymbol{J}(i)|=|E\left[ {{\boldsymbol{\tilde w}}\left( i \right){{\boldsymbol{z}}^T}\left( {{\boldsymbol{w}}\left( i \right)} \right)} \right]|\leq\frac{1}{\delta}|E\left[ {{\boldsymbol{\tilde w}}\left( i \right)}\boldsymbol{1}^T \right]|
\end{equation}
Therefore, to ensure the bounded property of $\boldsymbol{J}(i)$, $E\left[ {{\boldsymbol{\tilde w}}\left( i \right)} \right]$ should be always bounded. This recalls the mean recursion relation, which is obtained by taking expectation of both sides of Eq.(\ref{EMTC3})
\begin{equation}
\label{MeanR}
E[\tilde {\boldsymbol{w}}(i + 1)] = \boldsymbol{Q}E[\tilde {\boldsymbol{w}}(i)] - \xi {\boldsymbol{\cal A}}_2^T{\boldsymbol{{\cal D}}}E[{\boldsymbol{z}}\!\left( {{\boldsymbol{w}}\!\left( i \right)} \right)]
\end{equation}
where
\begin{equation}
\boldsymbol{Q}=\boldsymbol{\cal A}_2^T\left( {\boldsymbol{I} - \frac{1}{M}}\boldsymbol{\cal GD} \right)\boldsymbol{\cal A}_1^T
\end{equation}
It is known that $E\left[ {{\boldsymbol{\tilde w}}\left( i \right)} \right]$ will be bounded for all $i$ if $E\left[ {{\boldsymbol{\tilde w}}\left( i \right)} \right]$ converges as $i\rightarrow \infty$. Moreover, it has been proven that the stability of $\boldsymbol{\cal F}$ and $\boldsymbol{Q}$ can ensure the convergence of Eq.(\ref{EMTC5}) and Eq.(\ref{MeanR}), respectively \cite{Lorenzo2013Sparse}. Thus, for arbitrary $\xi$ the sufficient condition for both mean and mean square stability of Eq.(\ref{EMTC5}) should be
\begin{equation}
\label{rhoQ}
\rho(\boldsymbol{\cal F})<1~~and~~\rho(\boldsymbol{Q})<1
\end{equation}
where $\rho(\boldsymbol{X})$ denotes the spectral radius of matrix $\boldsymbol{X}$.
\par To further simplify the condition in Eq.(\ref{rhoQ}), we propose the following theorem:
\begin{myTheo}
\label{myTheo1}
For arbitrary real square matrices sequence $\{\boldsymbol{B}_k\}_{k=1}^{t}\in\boldsymbol{R}^{l\times l},t,l\in\boldsymbol{N}_+$, the following inequality will always hold
\begin{equation}
\rho\left(\boldsymbol{B}_1 \otimes \boldsymbol{B}_1 \right)\leq\rho\left(\sum_{k=1}^{t}\left(\boldsymbol{B}_k \otimes \boldsymbol{B}_k \right)\right)
\end{equation}
\end{myTheo}
\begin{pf}
See Appendix B.
\end{pf}
According to Theorem.\ref{myTheo1}, if we set $\boldsymbol{B}_1=\boldsymbol{Q}^T$ and $\boldsymbol{B}_{k}=\frac{\sqrt{N+1}}{M}{\boldsymbol{\cal A}}_1\boldsymbol{\cal T}_{k-1}{\boldsymbol{\cal D}}{{\boldsymbol{\cal A}}_2}$ for $k=2,...,P+1$, we can obtain
\begin{equation}
\rho(\boldsymbol{G}^T\otimes\boldsymbol{G}^T)\leq\rho(\boldsymbol{\cal F})
\end{equation}
Moreover, according to eigenvalue relationship of kronecker product, we can obtain $\rho(\boldsymbol{Q}^T\otimes\boldsymbol{Q}^T)=[\rho(\boldsymbol{Q})]^2$. Therefore, we have the following conditions
\begin{equation}
\rho(\boldsymbol{\cal F})<1\Longrightarrow\rho(\boldsymbol{Q}^T\otimes\boldsymbol{Q}^T)<1\Longrightarrow\rho(\boldsymbol{Q})<1
\end{equation}
Therefore, we can conclude that under Assumptions A1-A3, the condition $\rho(\boldsymbol{\cal F})<1$ will guarantee both mean and mean-square stability of the proposed algorithm.
\par For large scale data, computing the eigenvalue of $\boldsymbol{\cal F}$ is hard since the complexity grows significantly as $N$ increases. To simplify the computation of $\boldsymbol{\cal F}$, we further propose the following theorem:
\begin{myTheo}
Given sum of arbitrary real square matrices sequence $\{\boldsymbol{B}_k\}_{k=1}^{t}\in\mathbb{R}^{l\times l},t,l\in\boldsymbol{N}_+$
\begin{equation}
\boldsymbol{\cal B}=\sum_{k=1}^{t}\left(\boldsymbol{B}_k \otimes \boldsymbol{B}_k \right)
\end{equation}
By defining
\begin{equation}
{\boldsymbol{\cal B}}^{\otimes \boldsymbol{I}_N}=\sum_{k=1}^{t}\left(\left(\boldsymbol{B}_k \otimes \boldsymbol{I}_N\right) \otimes \left(\boldsymbol{B}_k \otimes \boldsymbol{I}_N\right) \right)
\end{equation}
where $\boldsymbol{I}_N$ is the arbitrary identity matrix. Then, we will have
\begin{equation}
\rho({\boldsymbol{\cal B}}^{\otimes \boldsymbol{I}_N})=\rho(\boldsymbol{\cal B})
\end{equation}
\end{myTheo}

\begin{pf}
See Appendix C.
\end{pf}

Rewriting Eq.(\ref{calF}) we can obtain
\begin{equation}
\begin{aligned}
\boldsymbol{\cal F}=& {\left[ {{{\boldsymbol{A}}_1}\!\left(\! {{\boldsymbol{I}} \!-\! \frac{1}{{ {M}}}{\boldsymbol{GD}}} \!\right)\!{{\boldsymbol{A}}_2} \!\otimes\! {\boldsymbol{I}}} \right] \!\otimes\! \left[ {{{\boldsymbol{A}}_1}\!\left(\! {{\boldsymbol{I}} \!-\! \frac{1}{{ {M}}}{\boldsymbol{GD}}} \!\right)\!{{\boldsymbol{A}}_2} \!\otimes\! {\boldsymbol{I}}} \right]}\\
&+\frac{{N + 1}}{{{M^2}}}\sum\limits_{k = 1}^M {\left[ {\left( {{{\boldsymbol{A}}_1}{{\boldsymbol{T}}_k}{\boldsymbol{D}}{{\boldsymbol{A}}_2} \otimes {\boldsymbol{I}}} \right) \otimes \left( {{{\boldsymbol{A}}_1}{{\boldsymbol{T}}_k}{\boldsymbol{D}}{{\boldsymbol{A}}_2} \otimes {\boldsymbol{I}}} \right)} \right]}
\end{aligned}
\end{equation}
Thus, according to Theorem.2, one can compute $\rho(\boldsymbol{\cal F})$ by simply set ${\boldsymbol{I}}=1$. Specifically, by defining
\begin{equation}
\begin{aligned}
\label{F}
\boldsymbol{F}=&\left( {{{\boldsymbol{ A}}_1} \otimes {{\boldsymbol{ A}}_1}} \right)\left[ {\left({\boldsymbol{I}} - \frac{1}{{{M}}}{\boldsymbol{ GD}} \right)\otimes \left({\boldsymbol{I}} - \frac{1}{{{M}}}{\boldsymbol{ GD}} \right)}\right.\\
&+ \left.{\frac{{N + 1}}{{{M^2}}}\sum\limits_{k = 1}^P {\left( {{{\boldsymbol{ T}}_k}{\boldsymbol{ D}} \otimes {{\boldsymbol{ T}}_k}{\boldsymbol{ D}}} \right)} } \right)\left( {{{\boldsymbol{ A}}_2} \otimes {{\boldsymbol{ A}}_2}} \right)
\end{aligned}
\end{equation}
we will have $\rho(\boldsymbol{F})=\rho(\boldsymbol{\cal F})$. Therefore, instead of calculating the spectrum radius of $\boldsymbol{\cal F}$ with $P^2N^2\times P^2N^2$ dimensions, we can simplify obtain $\rho(\boldsymbol{\cal F})$ from $\boldsymbol{F}$ which has only $P^2\times P^2$ dimensions. In practical, the $\boldsymbol{F}$ is typically a sparse matrix. Thus we can use an computationally-efficient search technique \cite{Douglas1995Exact,Golub2013Matrix} to find $\rho(\boldsymbol{F})$, which is easy to implement.\\


\subsection{Further analysis under general parameter settings}
Under Assumption A1-A3, $\rho(\boldsymbol{F})<1$ gives the sufficient condition for the convergence of the proposed algorithm. Moreover, in practical diffusion adaptation, the step size of all nodes are always set to the same value, and $\boldsymbol{S}$ is always set to doubly stochastic matrix \cite{Lopes2008Diffusion,Cattivelli2010Diffusion,Chen2012Diffusion,sayed2013diffusion,Abdolee2015Diffusion,Piggott2016Diffusion}. Therefore, in the following analysis we set $\boldsymbol{D}=\mu\boldsymbol{I}$ where $\mu$ is the identical step size for all nodes. The $\boldsymbol{S}$ is set to doubly stochastic such that $\boldsymbol{G}=\boldsymbol{I}$. Without loss of generality, here we only analysis ATC strategy, such that $\boldsymbol{A}_1=\boldsymbol{I}$. Thus Eq.(\ref{F}) can be further simplified as
\begin{equation}
\begin{aligned}
\label{Fs}
\boldsymbol{ F}=& \left[{\left(1 - \frac{\mu}{M} \right)^2 \boldsymbol{I}}+ {\frac{{N + 1}}{{{M^2}}}\mu^2\sum\limits_{k = 1}^M {\left( {{{\boldsymbol{ T}}_k} \otimes {{\boldsymbol{ T}}_k}} \right)} } \right]\left( {{{\boldsymbol{ A}}_2} \otimes {{\boldsymbol{ A}}_2}} \right)
\end{aligned}
\end{equation}
Under above parameter settings, one can further derive the following proposition.
\begin{myPro}
For arbitrary column stochastic matrix $\boldsymbol{A}_2$ and doubly stochastic matrix $\boldsymbol{S}$, the upper bound of the step size $\mu_{max}$ to guarantee $\rho(\boldsymbol{F})<1$ in Eq.(\ref{F}) will within the range
\begin{equation}
\label{mumax}
\frac{2M}{(N+1)\zeta+1}\leq{\mu_{max}}\leq\frac{2PM}{P+N+1}
\end{equation}
where $\zeta=\max\{{{\boldsymbol{S}}^T}{{\boldsymbol{S}}}\}$. Specifically, the maximum $\mu_{max}$ can be obtained when the network is fully connected with $\alpha_{l,k}=1/P$ for all $l$ and $k$.
\end{myPro}
\begin{pf}
See Appendix D.
\end{pf}
\par Proposition 2 reveals the step size improvement introduced by diffusion strategies. In particular, when $P=1$, $\mu_{max}$ will be $2M/(N+2)$, which coincides with the upper bound of $l_0$-LMS algorithm \cite{Jin2010A}. Moreover, when $P>1$, $\zeta\leq1$ will always hold, and $\mu_{max}$ will be always larger than $2M/(N+2)$. While when data scale $N$ is relatively large compared with network size $P$, $\mu_{max}$ can achieve nearly $P$ times of the step size upper bound of $l_0$-LMS. Since the formulation of D$l_0$-LMS and $l_0$-LMS are similar, it can be inferred that large step size will offer faster convergence speed. This fact will be confirmed by simulations in Section V.
\par Moreover, we can also search the exact $\mu_{max}$ based on Eq.(\ref{Fs}) under certain $\boldsymbol{A}_2$ and $\boldsymbol{S}$. Actually, experimental results show that $\rho(\boldsymbol{F})$ is a convex function of $\mu$ within the range $\mu\in[\frac{2M}{(N+1)\zeta+1},\frac{2PM}{P+N+1}]$. Thus, we can follow the numerical search algorithm proposed in \cite{Douglas1995Exact} to iteratively find the exact $\mu_{max}$ so that $\rho(\boldsymbol{F})=1$.

\subsection{Effect of regularization and data correlation on convergence condition}
\par In typical CS reconstruction task, apart form diffusion strategies, regularization parameter $\xi$ and data correlation also affect the $\mu_{max}$. When $\xi=0$, $\rho(\boldsymbol{F})<1$ is the necessary and sufficient condition for the convergence of Eq.(\ref{EMTC5}). While when $\xi\neq0$, we have known from analysis in \cite{Lorenzo2013Sparse} that $\rho(\boldsymbol{F})<1$ is only the sufficient condition for the convergence, so that $\xi$ will further increase the $\mu_{max}$. Moreover, to ensure the successful reconstruction of the sparse vector, one should select proper $\xi$ to constrain the sparsity of the estimation as well as achieve desirable accuracy. In practice, $\xi$ is always selected as small values, and the increment of $\mu_{max}$ by $\xi$ will be small.
\par On the one hand, the independent assumption A2 is hard to satisfy in practical due to limited number of data in each node. Therefore the effect of data correlation on convergence should be taken into consideration. Specifically, the step size upper bound $\mu_{max}$ of the adaptive filter under correlation input data has been analyzed in \cite{Douglas1995Exact}, which shows that the $\mu_{max}$ of LMS is more stringent than the bound predicted by the independence regressor assumption. Since the weight update form of each node is similar to LMS, we can deduce that the actual upper bound $\mu_{max}$ will also be less than that estimated under assumption A2. In particular, different from adaptive system analyzed in \cite{Douglas1995Exact} where the input data is pairwise related, the input data in CS is periodic related. When $L_k$ is large (i.e. each node has large number of data), the influence of correlation will be greatly reduced. While when $L_k$ is small, the effect of data correlation cannot be neglected, therefore $\mu_{max}$ will be distinctly smaller than theoretical estimation.

%

\section{Simulation}
\label{sec:pagestyle}

In this section we verify the performance of the proposed algorithm in CS reconstruction task. The locations of non-zero entries of the sparse vector $\boldsymbol{x}$ are randomly selected within $[1,N]$, and the corresponding values are independently generated from uniform distribution within $[-1,-0.2]\bigcup[0.2,1]$. Further, $\boldsymbol{x}$ is normalized to a unit vector. The Gaussian measurement matrix is used in the simulations, i.e. each entry of $\boldsymbol{\Theta}$ is generated from Gaussian distribution with zero mean and variance $1/M$. The noise $\boldsymbol{v}$ is zero mean Gaussian distributed with covariance matrix $\frac{{\sigma}^2}{M}\boldsymbol{I}_{M\times M}$. Then, the observed measurement $\boldsymbol{y}$ are obtained from Eq.(\ref{CS1}).
\par The reconstruction of a sparse vector $\boldsymbol{x}$ is carried out by a connected network with $P$ nodes. The measurement matrix $\boldsymbol{\Theta}$ and corresponding measurement $\boldsymbol{y}$ are equally assigned to each node so that $L_k\in\{\lfloor M/P\rfloor,\lfloor M/P\rfloor+1\}$, $k=1,2,..,P$. In combination step, the averaging weights are used so that ${\beta}_{l,k}=1/|{\cal N}_{k}|$ for all $l$. While in adaptation step, we use the Metropolis weights defined as
\begin{equation*}
\begin{aligned}
\renewcommand\arraystretch{1.1}
{\alpha_{l,k}} = \left\{
\begin{array}{lll}
\frac{1}{{\max \left\{ {{n_k},{n_l}} \right\}}},&l \in N_k\backslash \{k\} \\
1 - \sum\limits_{l \in N_k^ - }{\alpha_{l,k}} ,&l = k \\
0,&l \notin {N_k}
\end{array} \right.
\end{aligned}
\end{equation*}
Thus $\boldsymbol{S}$ is the doubly stochastic matrix.
\par The estimation misalignment at $i$-th iteration is evaluated by mean squared deviation (MSD) in dB, which is approximated as $10\log_{10}\{\frac{1}{T}\sum_{t=1}^{T}{{\|\frac{1}{P}\sum_{k=1}^P{{\boldsymbol{w}} _k}(i)-\boldsymbol{x} \|^2}}\}$ over $T$ numbers of Monte Carlo runs with different sparse signals, sensing matrices and noises. Further, for each Monte Carlo run, the reconstruction is considered successful if the average MSD value of last iteration is less than $1\times10^{-2}$.
During the whole simulation, without mentioned, the zero attraction parameters of the proposed algorithms are set as $\alpha=10$. The maximum iteration number $C=10^{5}$. The parameters in Eq.(\ref{Stop1}) and Eq.(\ref{Stop2}) are set as $\tau=1\times10^{-3}$, $p_s=20$ and $L_s=0.2N$.

\subsection{Convergence performance}
In this section, we investigate the convergence performance of the proposed algorithms along with $l_0$-LMS. The simulation is carried by a connected network with 20 nodes. Each node is linked to 3 nodes in average. The CS system parameters are set as $N=20000$, $M=4000$ and $K=500$. The noise variance $\sigma$ is set to $3\times 10^{-3}$. The regularization parameter $\xi$ for all algorithms are set to $5\times 10^{-8}$. The mini batch size $Q$ is set to 5. The step sizes are set as $0.4$, $4$, $16$ for $l_0$-LMS, gradient descend diffusion algorithms (ATC-D$l_0$-LMS and CTA-D$l_0$-LMS) and mini-batch diffusion algorithms (ATC-MB-D$l_0$-LMS and CTA-MB-D$l_0$-LMS), respectively. Note that when $\mu=4$ and $16$, traditional $l_0$-LMS and will diverge according to the analysis in \cite{Jin2010A}. While ATC-D$l_0$-LMS and CTA-D$l_0$-LMS will diverge when $\mu=16$ according to the analysis in Section IV.


\begin{figure}[tb]
\centering
\subfigure[]
{
\includegraphics[width=0.94\linewidth]{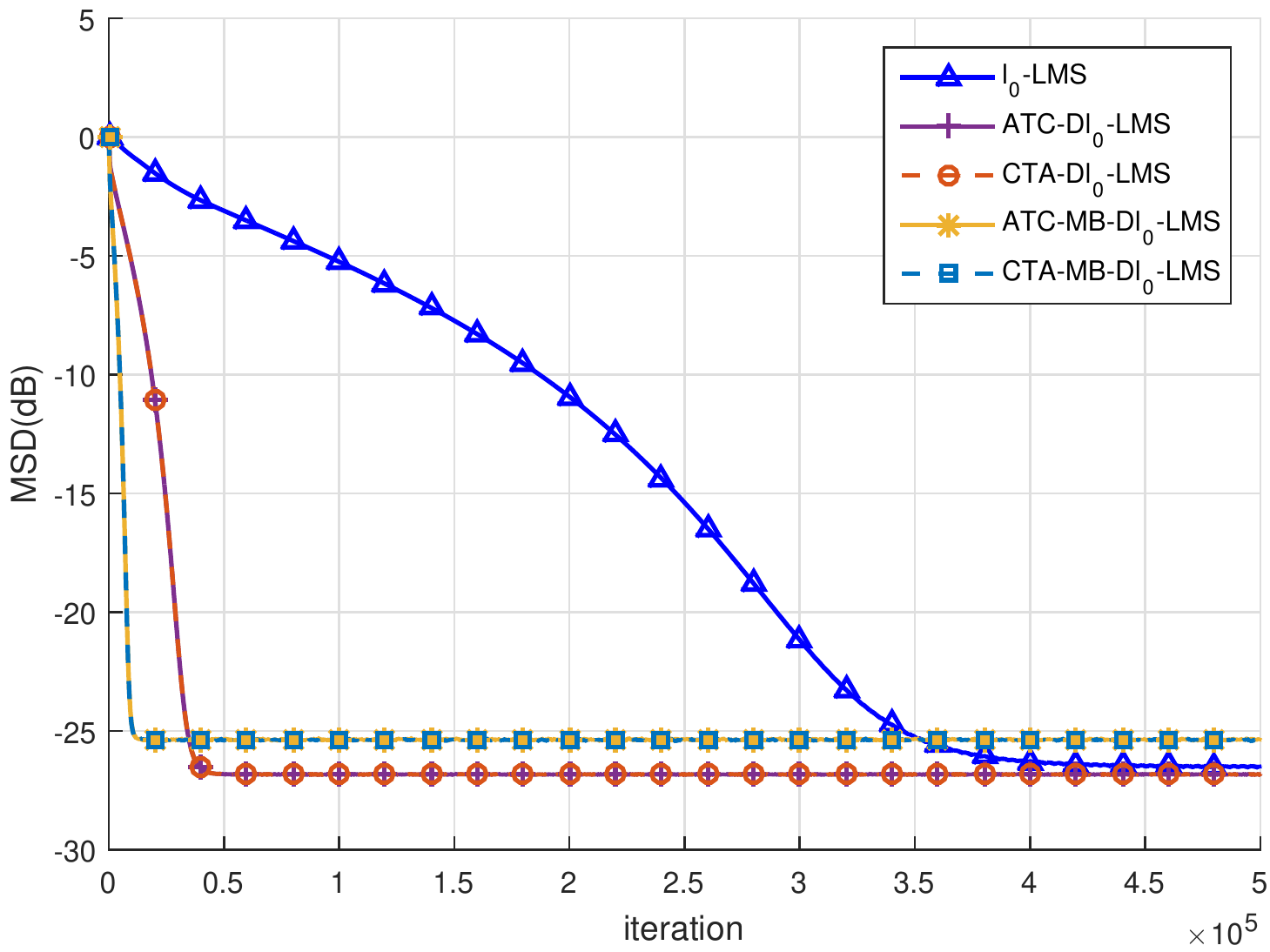}\hspace{0.15in}}
\hspace{0.1in}
\subfigure[]{
\includegraphics[width=0.94\linewidth]{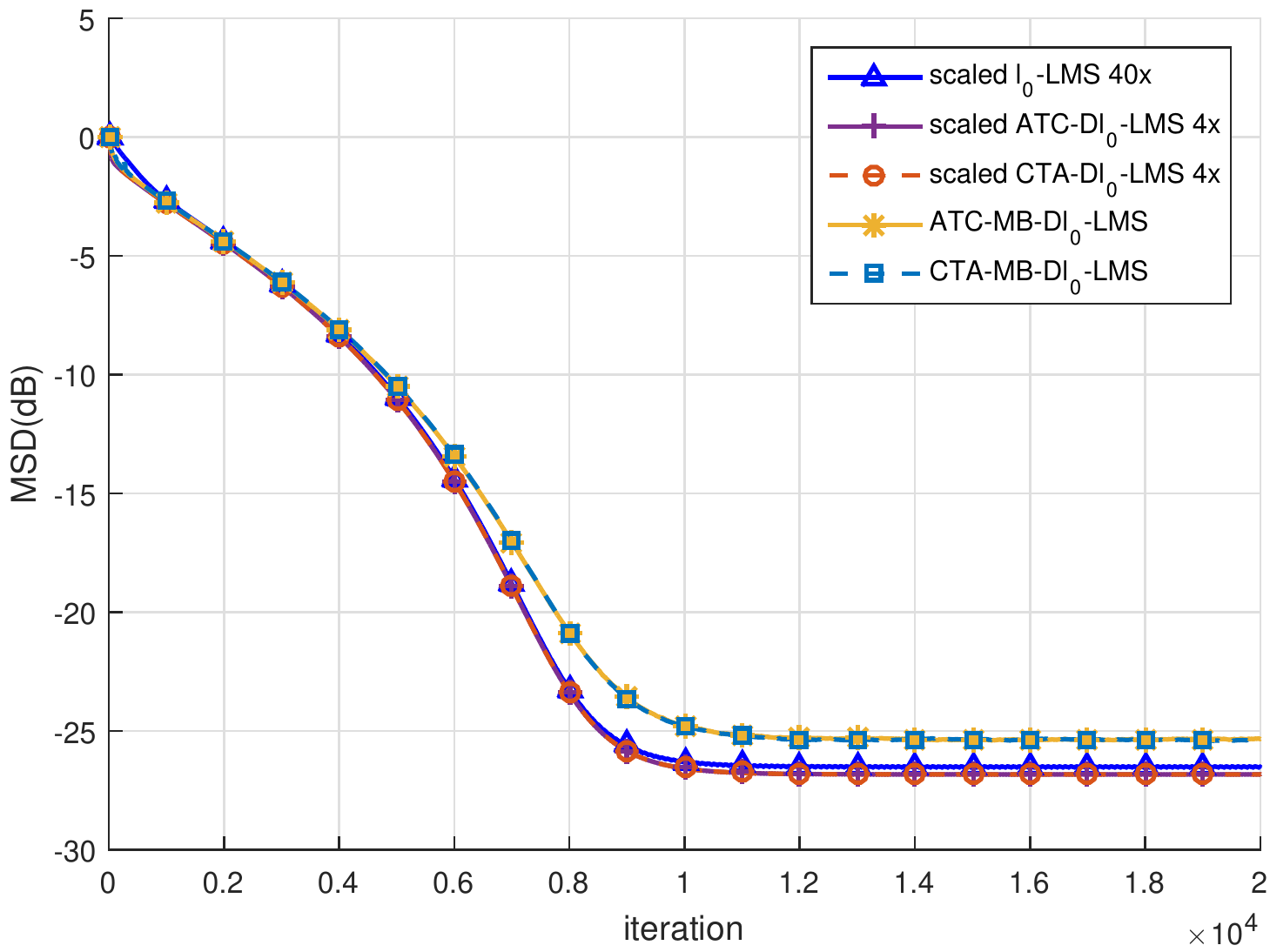}\hspace{0.15in}}
\caption{Average learning curves of different algorithms. (a) Original average learning curves of different algorithms in terms of iteration $i$.  (b) Scaled average learning curves of different algorithms. The learning curves of ATC-D$l_0$-LMS and D-$l_0$-LMS-CTA shrink 4 times, while the learning curve of $l_0$-LMS shrinks 40 times.}
\label{convergence}
\end{figure}

%
%

The average learning curves of all algorithms are shown in Fig.\ref{convergence}(a). 
Note that the stop criterion is not used in this simulation. One can observe that ATC-D$l_0$-LMS and D-$l_0$-LMS-CTA achieves similar reconstruction MSD with traditional $l_0$-LMS, while converge much faster than $l_0$-LMS. ATC-MB-D$l_0$-LMS and MB-D-$l_0$-LMS-CTA can further converge faster than ATC-D$l_0$-LMS and CTA-D$l_0$-LMS with a slight loss on reconstruction accuracy. Further, one can also observe that the convergence behaviour of ATC strategy is quite similar with CTA.
\par To further investigate the relation between step size and convergence speed, we scale down the learning curves in Fig.\ref{convergence}(a). The scale factors of each algorithm are obtained according to the step sizes. The scaled learning curves are shown in Fig.\ref{convergence}(b). One can observe that the scaled learning curves are similar. The results confirm that the convergence speed is closely related to the step size under the same regularization parameter $\xi$.

\begin{figure*}[tb]
\centering
\subfigure
[]{
\includegraphics[width=0.45\linewidth]{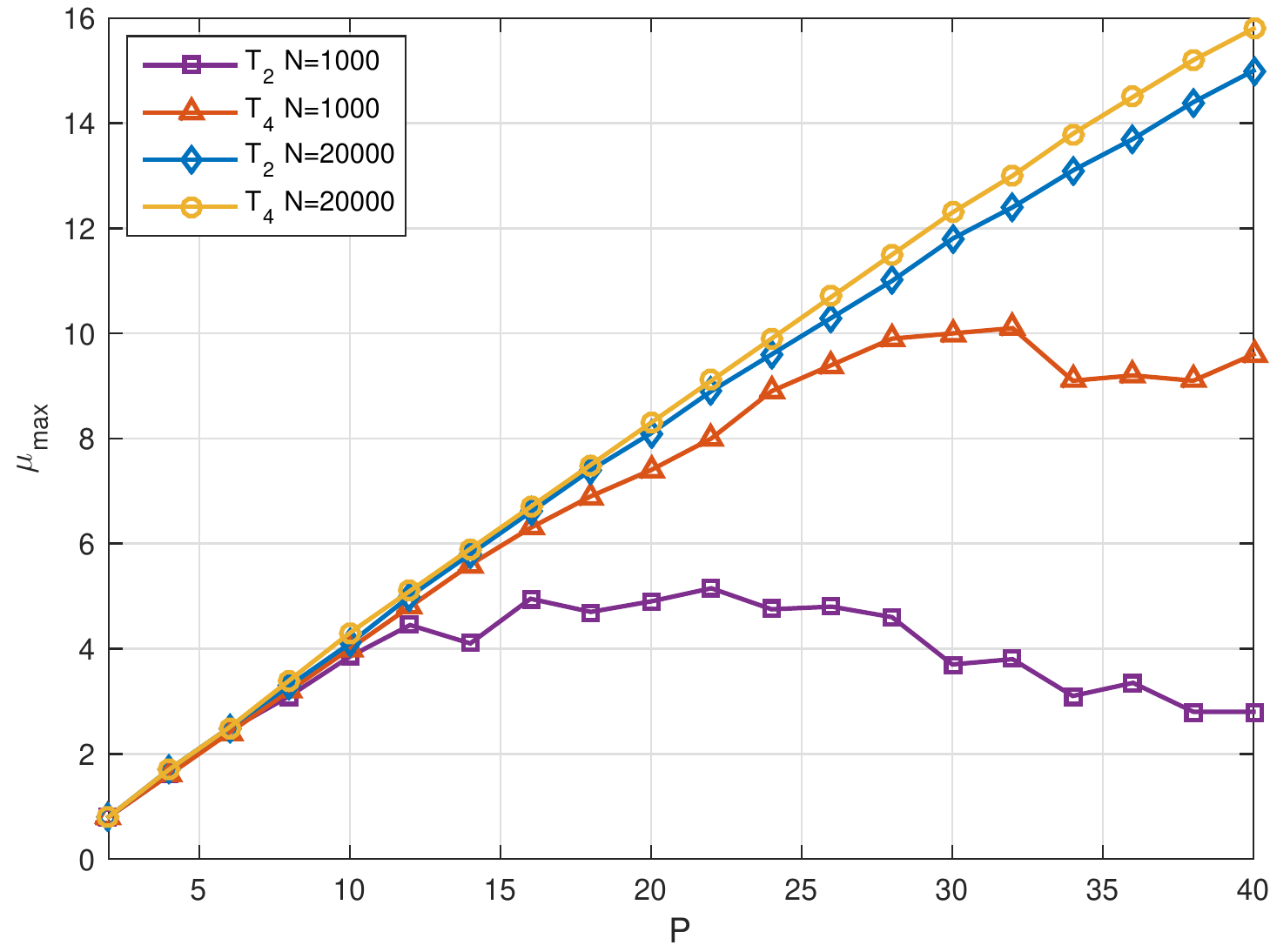}\hspace{0.15in}}
\hspace{0.1in}
\subfigure[]{
\includegraphics[width=0.45\linewidth]{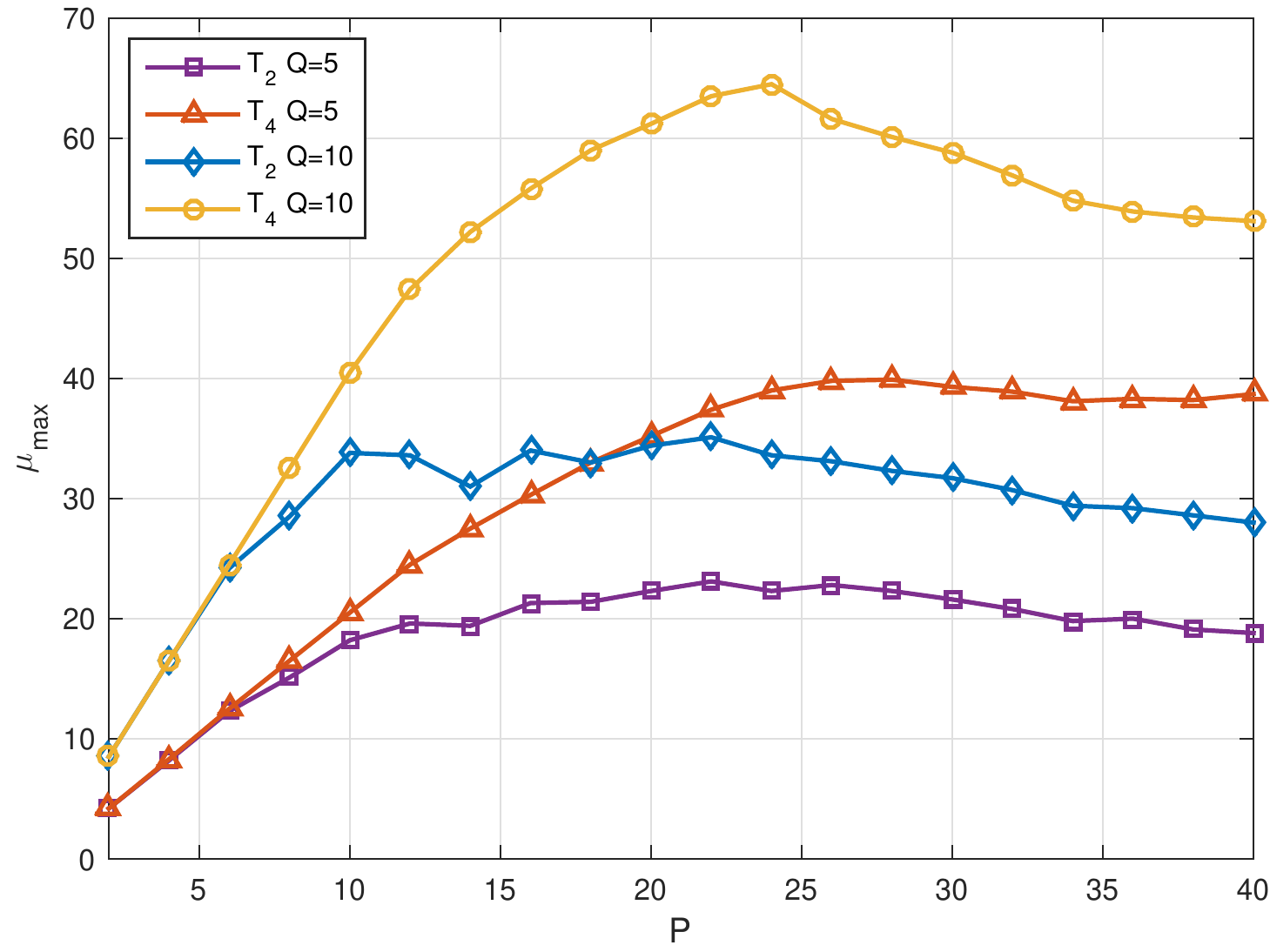}\hspace{0.15in}}
\subfigure[]{
\includegraphics[width=0.45\linewidth]{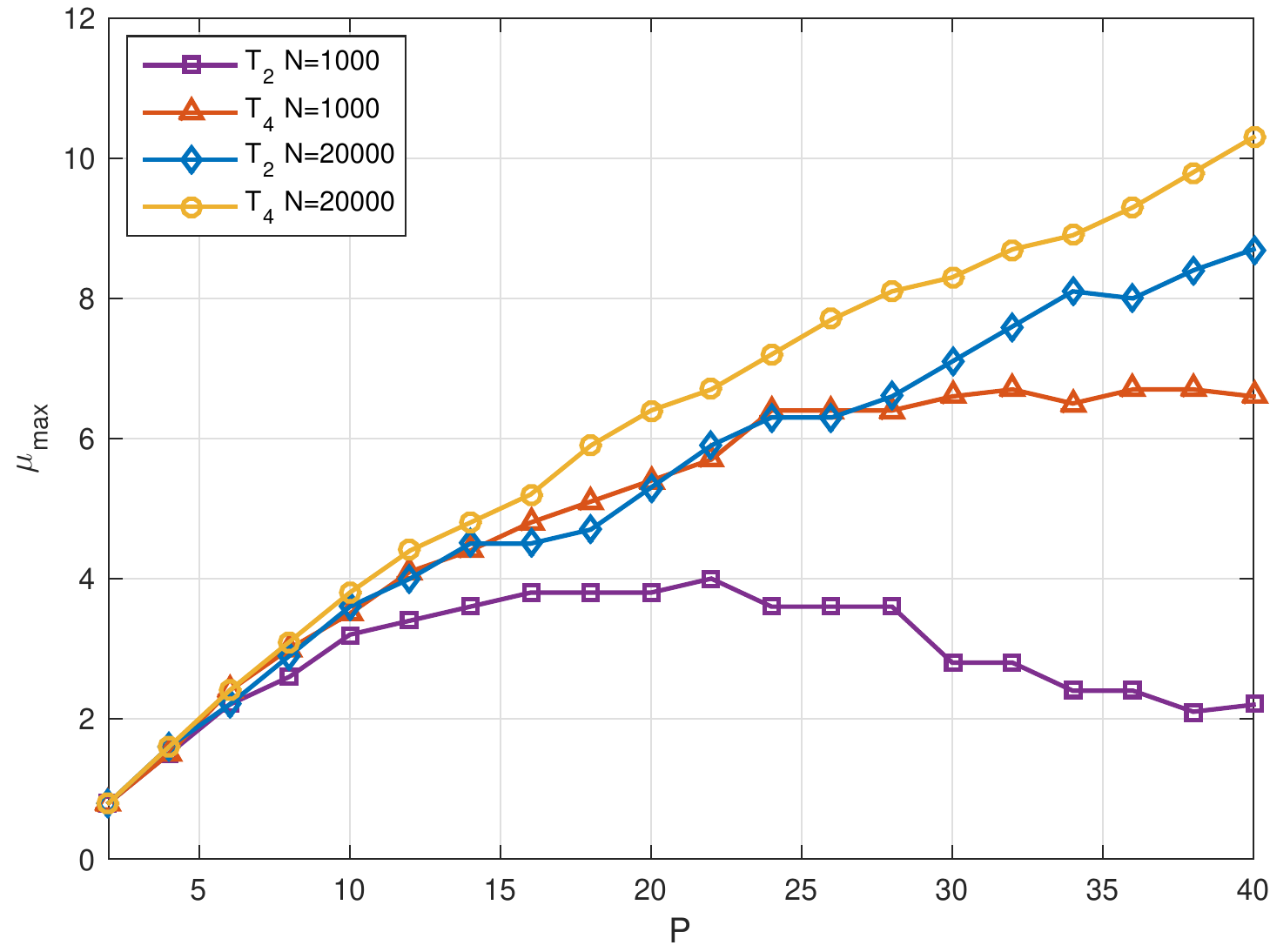}\hspace{0.15in}}
\hspace{0.1in}
\subfigure[]{
\includegraphics[width=0.45\linewidth]{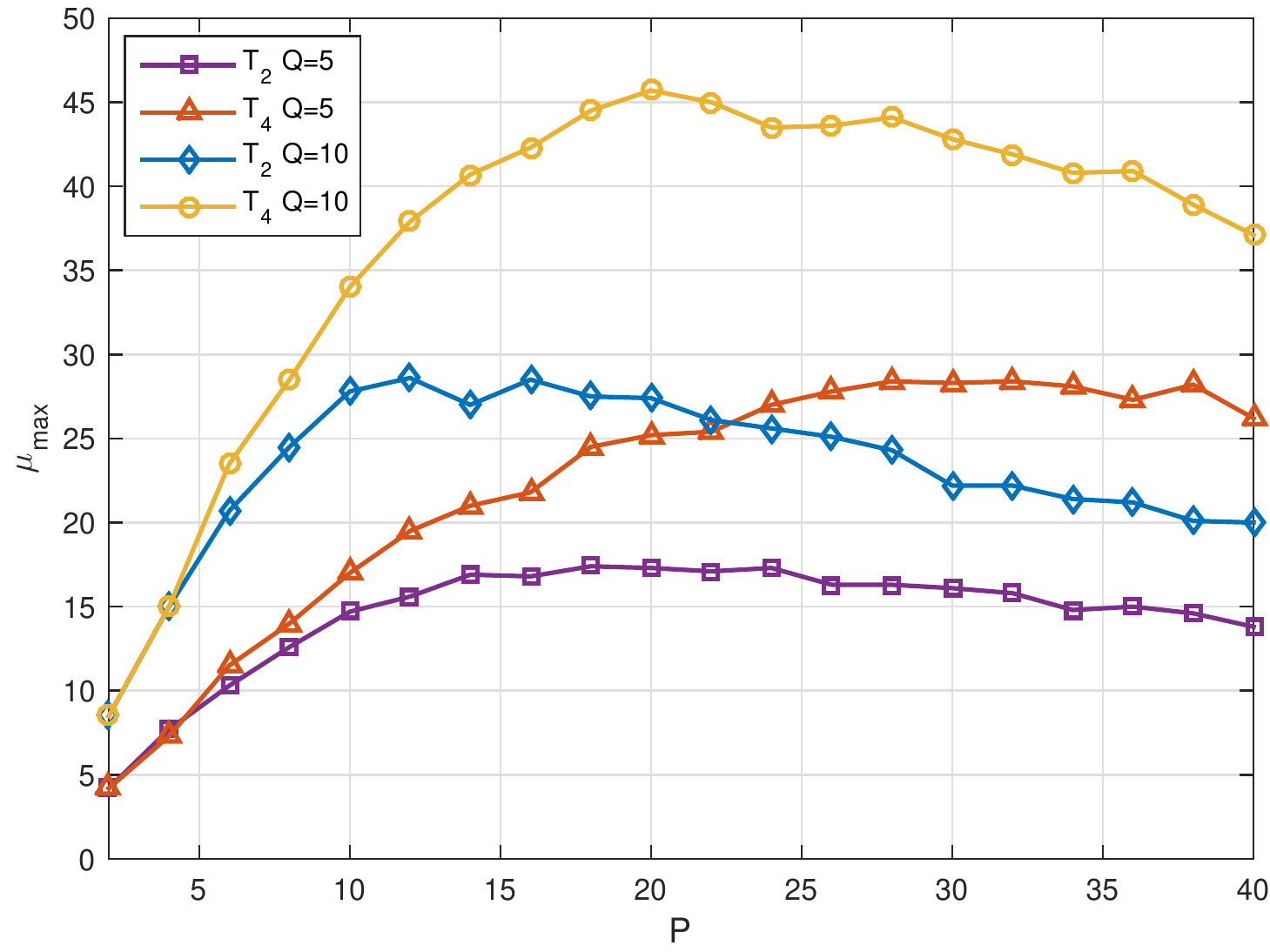}\hspace{0.15in}}
\caption{The step size upper bound $\mu_{max}$ under different network size $P$. (a) ATC-D$l_0$-LMS with $p=2,4$ and $N=1000,20000$. (b) ATC-MB-D$l_0$-LMS with $p=2,4$ , $Q=5,10$ and $N=20000$. (c) ATC-D$l_0$-LMS without adaptation information exchange (i.e.$\boldsymbol{S}$=$\boldsymbol{I}$) with $p=2,4$ and $N=1000,20000$. (d) ATC-MB-D$l_0$-LMS without adaptation information exchange (i.e.$\boldsymbol{S}$=$\boldsymbol{I}$) with $p=2,4$ , $Q=5,10$ and $N=20000$.}
\label{fig5}
\end{figure*}

\subsection{Step size upper bound}
In this part, we investigate the relationship between step size upper bound $\mu_{max}$ and network size $P$. {\color{black}The network size is gradually increased from 2 to 40. To ensure the continuity of the growth of the network, 
we recursively generated the network by linking a new node to $p$ nodes of the previous network, and 
the generated network set with $p$ is represented as $T_{p}$. The parameter $p$ actually controls the average link numbers of each node, a larger $p$ indicates more information communications across the network at each iteration. } In the following simulations, $\mu_{max}$ is computed as the largest step size that can reconstruct all the sparse signals under 200 Monte carlo runs.
\par First, we investigate the $\mu_{max}$ of ATC-D$l_0$-LMS under different scale of data. The dimension $N$ of the sparse vector is set to $1000$ and $20000$. The number of measurement $M$ is set to $0.2N$, and the sparsity $K$ is set to $0.125M$. The regularization parameter $\xi$ for all algorithms are set as $5\times 10^{-6}$ for $N=1000$ and $5\times 10^{-8}$ for $N=20000$. Fig.\ref{fig5}(a) depicts the curves of network size $P$ versus $\mu_{max}$ under different network set $T_2$ and $T_4$. As can be seen, when $N=20000$, $\mu_{max}$ can grows linearly under $T_2$ and $T_4$. While for $N=1000$, the growth of $\mu_{max}$ is limited when $P$ is large.

\begin{figure*}[tb]
\centering
\includegraphics[width=0.99\linewidth]{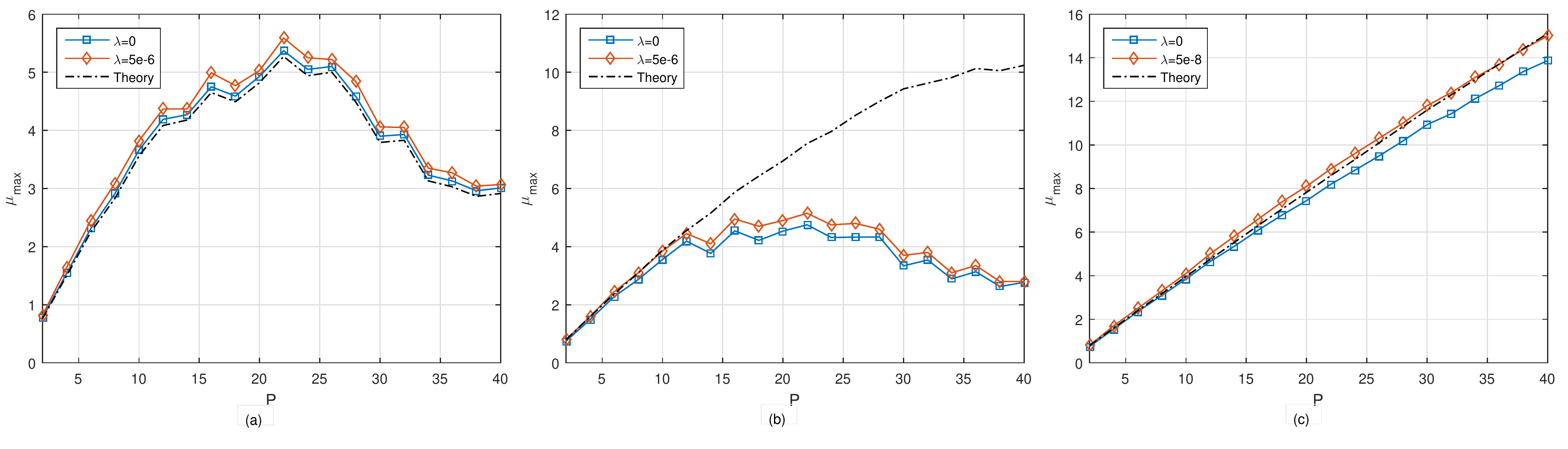}
\caption{Simulated and theoretical $\mu_{max}$ in terms of Fig.\ref{fig5}(a) under $p=2$. (a) Data dimension $N=1000$. The theoretical $\mu_{max}$ is obtained using Eq.(\ref{Gamma}) and Eq.(\ref{T1}). (b) Data dimension $N=1000$. The theoretical $\mu_{max}$ is obtained using Eq.(\ref{F}). (c) Data dimension $N=20000$. The theoretical $\mu_{max}$ is obtained using Eq.(\ref{F}). }
\label{theory}
\end{figure*}

{\color{black}We also verify the theoretical analysis with above simulation. The theoretical $\mu_{max}$ corresponds to simulated $\mu_{max}$ with $\xi=0$. Note that the reconstruction can not succeed without $\xi$, thus the simulated $\mu_{max}$ with $\xi=0$ is calculated as the largest step size that ensure the algorithms not diverge. In Fig.\ref{theory}(a), the simulation is carried out with $N=1000$ and the theoretical results are obtained using Proposition 1 (Eq.(\ref{Gamma}) and Eq.(\ref{T1})). The measurement matrix $\boldsymbol{\Theta}$ is fixed during Monte carlo runs. One can see that the theoretical $\mu_{max}$ match well with simulation results. Then, the theoretical results computed from Eq.(\ref{F}) versus simulation results are given in Fig.\ref{theory}(b)-(c). One can observe that for $N=20000$, the simulated and theoretical $\mu_{max}$ are similar. While for $N=1000$, due to strong correlation of data, the simulated $\mu_{max}$ are much lower than theoretical analysis when network size is large.} Moreover, one can also observe from Fig.\ref{theory}(a)-(c) that the curves of simulated $\mu_{max}$ with proper $\xi$ are always above the curves without $\xi$, which confirms the step size upper bound improvement by $\xi$.  

\par Second, the $\mu_{max}$ of ATC-MB-D$l_0$-LMS under $N=20000$ is performed. The curves of network size $P$ versus $\mu_{max}$ under different parameter $p$ and batch size $Q$ are shown in Fig.\ref{fig5}(b). One can see that $\mu_{max}$ is further greatly improved compared with ATC-D$l_0$-LMS. Specifically, a larger batch size $Q$ and link parameter $p$ allow larger $\mu_{max}$. Moreover, the results show that mini-batch method gives remarkable acceleration for small network, while the growth of $\mu_{max}$ is limited when network is large.
\par We also investigate the performance of ATC-D$l_0$-LMS and ATC-MB-D$l_0$-LMS without adaptation information exchange. All the parameters are the same as the first simulation in this part except $\boldsymbol{S}=\boldsymbol{I}$. Fig.\ref{fig5}(c) and Fig.\ref{fig5}(d) show the curves of $\mu_{max}$ versus network size $P$ for ATC-D$l_0$-LMS and ATC-MB-D$l_0$-LMS, respectively. One can observe that without adaptation step, $\mu_{max}$ suffers from different degrees of decline. Nevertheless, the acceleration is still significant compared with $l_0$-LMS. In general, dropping away the adaptation information exchange reduces $50\%$ and $66.7\%$ of the network data transmission for D-$l_0$-LMS and MB-D-$l_0$-LMS, respectively.
\par
\begin{figure}[tb]
\centering
\includegraphics[width=0.9\linewidth]{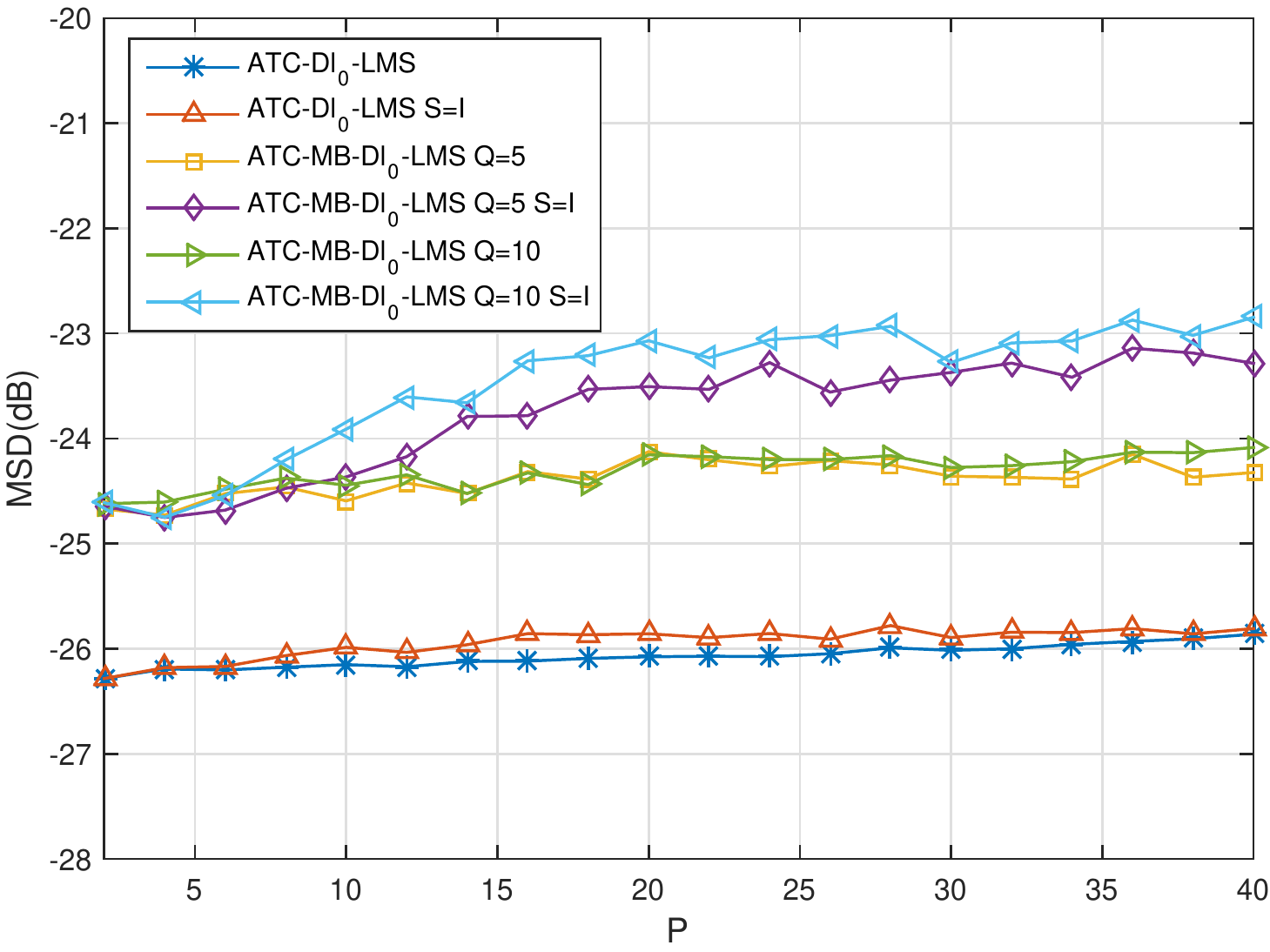}
\caption{Reconstruction MSD in terms of Fig.\ref{fig5} under $p=2$ and $N=20000$}
\label{N20000_Smax_MSD}
\end{figure}

 Finally, the reconstruction MSD values under different network sizes are conducted. Fig.\ref{N20000_Smax_MSD} shows the reconstruction MSD with $\mu_{max}$ from Fig.\ref{fig5}(a)-(d) under $T_2$ and $N=20000$. One can observe that the different network sizes $P$ affect slightly on reconstruction MSD except ATC-MB-D$l_0$-LMS without adaptation information exchange. Specifically, when $P$ is large, ATC-MB-D$l_0$-LMS has the performance loss of about 1.5dB when dropping away adaptation information exchange. Further, ATC-D$l_0$-LMS achieves lower reconstruction MSD compared with ATC-MB-D$l_0$-LMS.

%


\subsection{Sensitivity of regularization parameter $\xi$ and noise variance $\sigma$}
\par
\begin{figure}[tb]
\centering
\includegraphics[width=0.94\linewidth]{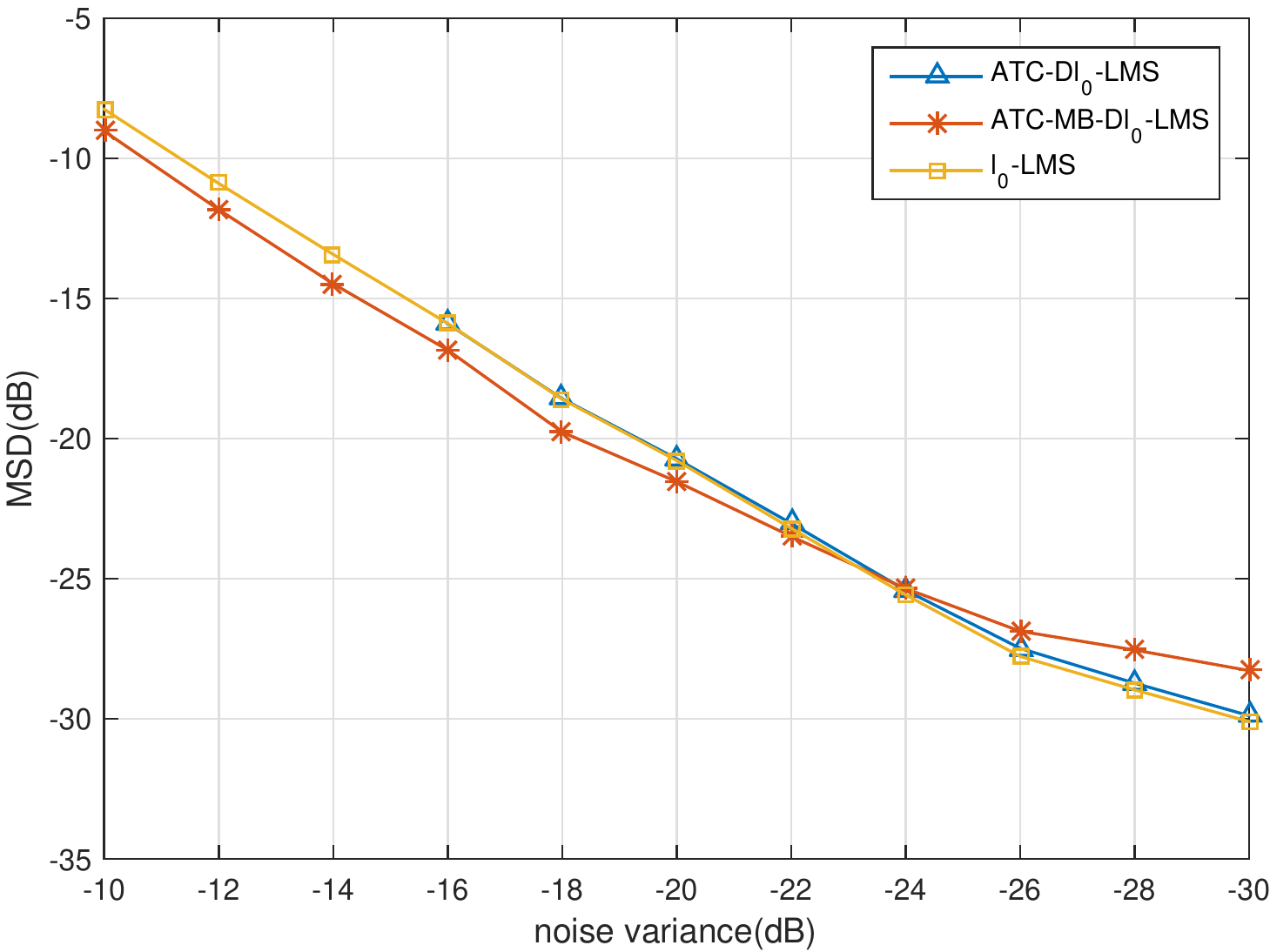}
\caption{Reconstruction MSD under different noise variance $\sigma$.}
\label{noise}
\end{figure}

 In this part we first focus on the reconstruction MSD under different noise variance $\sigma$. We select Gaussian noise with different variances $\sigma$ as the additive noise of the CS system. The simulation settings are the same as in part $A$. For each $\sigma$, 200 Monte Carlo runs are performed. Fig.\ref{noise} depicts the average MSD curves of three algorithms under different $\sigma$. One can see that ATC-D$l_0$-LMS and $l_0$-LMS achieve similar reconstruction performance in different noise situations. ATC-MB-D$l_0$-LMS achieves slightly better performance than ATC-D$l_0$-LMS and $l_0$-LMS when noise variance is larger than -24dB, while the performance is slightly worse when noise is less than -24dB.

\par
\begin{figure}[tb]
\centering
\includegraphics[width=0.94\linewidth]{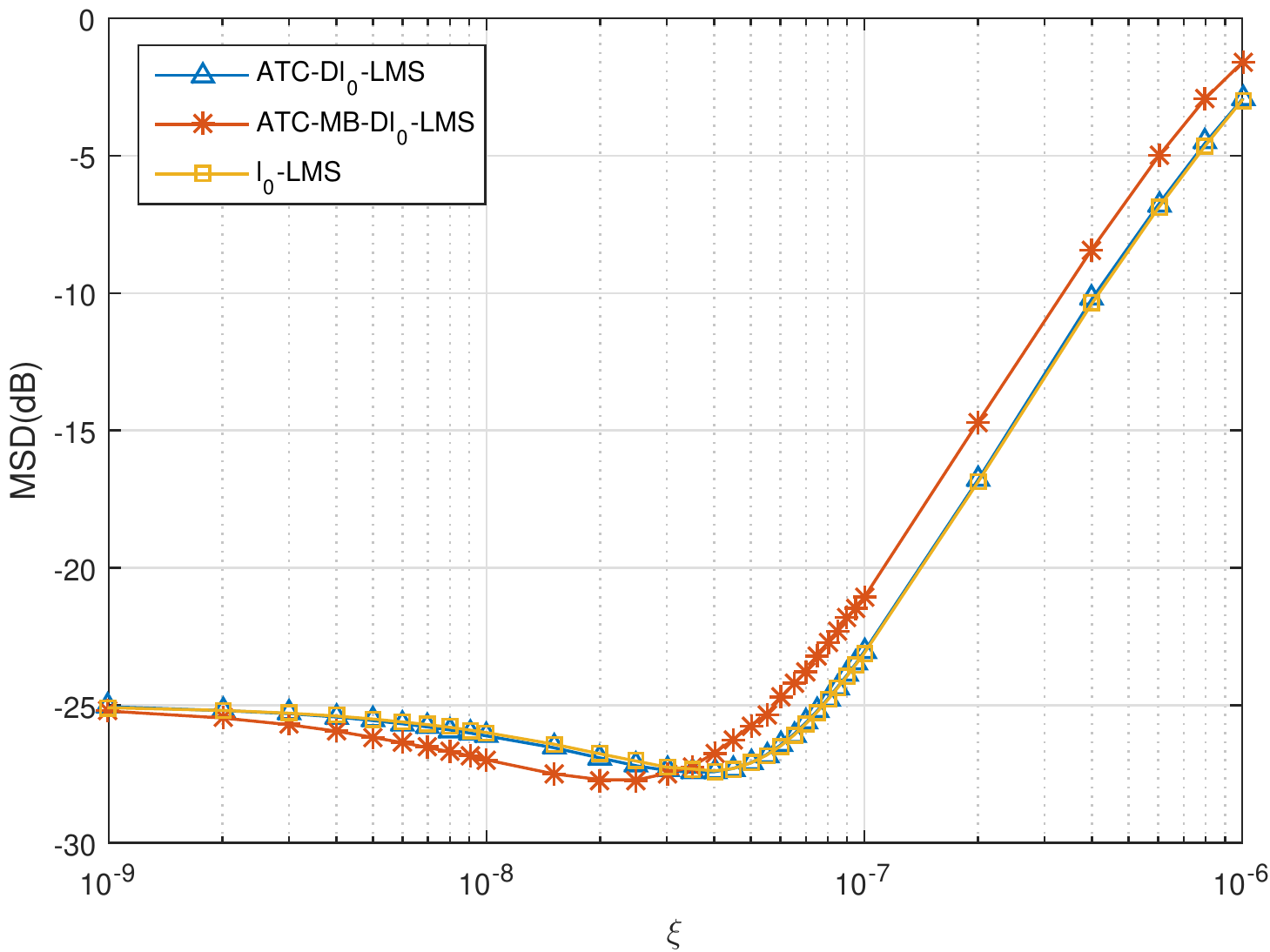}
\caption{Reconstruction MSD under different regularization parameter $\xi$.}
\label{lambda}
\end{figure}

 We further compare the reconstruction performance under different regularization parameter $\xi$. The simulation settings are the same as in part $A$. The maximum iteration number $C$ is not used in the simulation, so that all the reconstruction MSD values are obtained at steady state. For each $\xi$, 200 Monte Carlo runs are performed. The curves of average MSD versus $\xi$ are shown in Fig.\ref{lambda}. One can see that all algorithms can achieve good reconstruction performance when $\xi$ is not too large (i.e. $\xi\leq10^{-7}$). Further, the curve of ATC-D$l_0$-LMS and $l_0$-LMS are quite similar, while the curve of ATC-MB-D$l_0$-LMS is moved from the curve of $l_0$-LMS about $2\times10^{-8}$ to the left. The results show that the proposed algorithms can achieve similar reconstruction MSD with $l_0$-LMS. For ATC-MB-D$l_0$-LMS, one can adjust a slightly lower $\xi$ to obtain the similar performance with ATC-D$l_0$-LMS and $l_0$-LMS. Moreover, although the algorithms work well for a wide range of $\xi$, one should note that smaller $\xi$ will cause slow convergence speed. One should select $\xi$ in a proper range to make trade off between reconstruction accuracy and convergence speed.

\section{Conclusion}

In this paper we propose a novel diffusion adaptation framework for CS reconstruction task. Using distributed network, the measurement matrix can be stored in a decentralized manner, thus the storage in each node can be efficiently reduced. Based on diffusion adaptation strategy, the gradient-descend diffusion CS reconstruction algorithm called D$l_0$-LMS is proposed. D$l_0$-LMS can collaboratively estimate the sparsity and recover the sparse signal across the network. Particularly, the convergence of D$l_0$-LMS is analyzed. To further improve the convergence speed, a mini-batch based diffusion algorithm is also proposed. Simulation results confirm the desirable performance of the proposed algorithms. Compared with $l_0$-LMS algorithm, the proposed algorithms can achieve much faster convergence speed while obtaining similar reconstruction accuracy.

\section*{Acknowledgements}
This work was supported by 973 Program (No.2015CB351703), National NSF of China (No.61273366 and No.91648208) and National Science and Technology support program of China (No.2015BAH31F01).

\begin{appendices}
\subsection{Appendix A}
{\color{black}First we prove that for given $\boldsymbol{\Theta}$, $\boldsymbol{A}_1$, $\boldsymbol{S}$, $\boldsymbol{A}_2$, for arbitrary selection of $\mu_{k}$, there must have $N-M$ eigenvalues of $\boldsymbol{\Gamma}$ with values equal to 1. For further analysis, we rewrite $\boldsymbol{\Gamma}$ as the block matrix with $P\times P$ blocks $\{\boldsymbol{\Gamma}_{ij}\}\in{\mathbb{R}}^{N\times N}$.

We also observe that $ {{\boldsymbol{I}} - {\boldsymbol{\cal DH}}\left( n \right)}$ is the block diagonal and Hermitian matrix with $P\times P$ blocks, 
and $\boldsymbol{\cal A}_1$, $\boldsymbol{\cal A}_2$ are also block matrices with the entries $\boldsymbol{\cal A}_{1,ij}=\beta_{i,j}\boldsymbol{I}$, $\boldsymbol{\cal A}_{2,ij}=\gamma_{i,j}\boldsymbol{I}$ respectively. Suppose there exists a vector $\boldsymbol{x}$ so that ${\boldsymbol{u}_k(n)}^T\boldsymbol{x}=0$ for arbitrary $n\in\{1,..L_m\}$ and $k\in\{1,...,P\}$, we can obtain
\begin{equation}
\boldsymbol{\Gamma}^T_{ij}\boldsymbol{x}=c_{ij}\boldsymbol{x}
\end{equation}
where $c_{ij}$ is the $\{i,j\}$-th entry of matrix $\boldsymbol{C}=(\boldsymbol{A}_1^T\boldsymbol{A}_2)^{L_m}$. Thus, by computing the non-vanishing vector $\boldsymbol{b}$ that satisfies $\left(\boldsymbol{C}-\boldsymbol{I}\right)\boldsymbol{b}=\boldsymbol{0}$ and constructing the column vector $\boldsymbol{\chi}={\rm{col}}\{b_k{\boldsymbol{x}}\}_{k=1}^{P}$, one can finally obtain
\begin{equation}
\label{chi}
\boldsymbol{\Gamma}\boldsymbol{\chi}=\boldsymbol{\chi}
\end{equation}
That is, $\boldsymbol{\chi}$ is the eigenvector of $\boldsymbol{\Gamma}$ with the corresponding eigenvalue 1.

The left problem is to find the vector $\boldsymbol{x}$ that satisfies the condition. Denoting $\{\boldsymbol{x}_i\}_{i=1}^S$ as a set of linearly independent vectors which are orthogonal to all rows of $\boldsymbol{\Theta}$, i.e.
\begin{equation}
\boldsymbol{\Theta}\boldsymbol{x}_i=0,~i=1,..,S
\end{equation}
Since $\boldsymbol{\Theta}$ is full row rank with rank $M$, we can derive that $S=N-M$. Therefore, there are $N-M$ numbers of vectors $\boldsymbol{x}$ ($\boldsymbol{\chi}$) that satisfy Eq.(\ref{chi}) and the conclusion is proved.

Based on the above conclusion, we further analyze the convergence of Eq.(\ref{EMTC3}) without regularization term. Setting $\xi=0$, and using recursion form in Eq.(\ref{EMTC3}), we can obtain
\begin{equation}
\label{recursion1}
\begin{aligned}
\!\tilde {\boldsymbol{w}}(i \!+\! 1)\! =\! &\prod_{n=1}^i\boldsymbol{\cal B}(n)\tilde {\boldsymbol{w}}(1) 
\!-\! \sum_{n=0}^{i-1}\left[\prod_{m=0}^{n-1}\boldsymbol{\cal B}(i\!-\!m)\boldsymbol{\cal A}_2^T\boldsymbol{\cal D}\boldsymbol{s}(i\!-\!n)\right]
\end{aligned}
\end{equation}
Where $\boldsymbol{\cal B}(n)={{\boldsymbol{\cal A}}_2^T\left[ {{\boldsymbol{I}} - {\boldsymbol{\cal DH}}\left( n \right)} \right]{{\boldsymbol{\cal A}}_1}}$. For simplicity, without mentioned, we define $\prod_{m=0}^{-1}\boldsymbol{X}=\boldsymbol{I}$ for arbitrary $\boldsymbol{X}$. Consider the fact that $\boldsymbol{s}(i)$ is also recursively used with period $L_m$, we write Eq.(\ref{recursion1}) at $(i\times L_m+1)$-th iteration as
\begin{equation}
\begin{aligned}
\label{recursion2}
\tilde {\boldsymbol{w}}(i\!\times\! L_m \!&+\! 1)\! = (\boldsymbol{\Gamma^T})^i\tilde {\boldsymbol{w}}(1)\\ 
&\!- \sum_{n=0}^{i-1}(\boldsymbol{\Gamma^T})^n\!\sum_{m=1}^{L_m}\!\left[\prod_{k=1}^{L_m\!-\!m}\!\!\boldsymbol{\cal B}(L_m\!-\!k\!+\!1)\boldsymbol{\cal A}_2^T\boldsymbol{\cal D}\boldsymbol{s}(m)\right]
\end{aligned}
\end{equation}
It is easy to conclude that the convergence for all $i$ in Eq.(\ref{recursion2}) can ensure the convergence for all $i$ in Eq.(\ref{recursion1}). To ensure the convergence of Eq.(\ref{recursion2}), one should make sure that both terms on the right side of Eq.(\ref{recursion2}) converges as $i$ tends to infinity. Using eigendecomposition we obtain $\boldsymbol{\Gamma}=\boldsymbol{WZ}\boldsymbol{W}^{-1}$ where each column of $\boldsymbol{W}$ is composed by eigen vectors of $\boldsymbol{\Gamma}$ and $\boldsymbol{Z}$ is a diagonal matrix with entries composed by eigenvalues of $\boldsymbol{\Gamma}$. Then $\boldsymbol{\Gamma}^i=\boldsymbol{W}\boldsymbol{Z}^i\boldsymbol{W}^{-1}$. Therefore, to guarantee the convergence of $\boldsymbol{\Gamma}^i$, the modulus of diagonal entries of $\boldsymbol{Z}$ should be no more than 1. From Eq.(\ref{chi}) one can obtain that $\boldsymbol{Z}$ consists at least $N-M$ diagonal entries with value 1. Thus to ensure the convergence of $\boldsymbol{\Gamma}^i$, $|\lambda_1|$ must be no more than 1.

To ensure the convergence of the second term on the right side of Eq.(\ref{recursion2}), we should prove
\begin{equation}
\label{recursion3}
(\boldsymbol{\Gamma}^T)^n\sum_{m=1}^{L_m}\!\left[\prod_{k=1}^{L_m\!-\!m}\!\!\boldsymbol{\cal B}(L_m\!-\!k\!+\!1)\boldsymbol{\cal A}_2^T\boldsymbol{\cal D}\boldsymbol{s}(m)\right]\rightarrow 0,n\rightarrow \infty
\end{equation}
Then, replacing $(\boldsymbol{\Gamma}^T)^n$ by $\boldsymbol{W}^{-T}\boldsymbol{Z}^n\boldsymbol{W}^{T}$, and using the fact that $\boldsymbol{x}^T{\boldsymbol{u}_k(n)}=0$, one can obtain that Eq.(\ref{recursion3}) will satisfy only if $\{|\lambda_i|\}_{i=N-M+1}^{N\times P}$ are all less than one. Therefore, based on the above analysis, the convergence of Eq.(\ref{recursion2}) will be guaranteed if $|\lambda_{N-M+1}|<1$.

Finally, we discuss the convergence with the regularization term ${\boldsymbol{z}_{\delta}}\!\left( {{\boldsymbol{w}}\!\left( i \right)} \right)$. According to the definition in Eq.(\ref{ZA}), we can obtain ${\boldsymbol{z}_{\delta}}\!\left( {{\boldsymbol{w}}\!\left( i \right)} \right)=\boldsymbol{0}$ when all the absolute values of elements of ${{\boldsymbol{w}}\!\left( i \right)}$ are large than $1/\delta$. If adding ${\boldsymbol{z}_{\delta}}\!\left( {{\boldsymbol{w}}\!\left( i \right)} \right)$ causes the divergence, ${\boldsymbol{z}_{\delta}}\!\left( {{\boldsymbol{w}}\!\left( i \right)} \right)$ will tend to zero vector and the recursion becomes equal to Eq.(\ref{recursion1}). Therefore, the convergence condition for Eq.(\ref{recursion1}) is still applicable for Eq.(\ref{EMTC3}) with regularization term.}

\subsection{Appendix B}
We follow the proof in \cite{liu1992spectral} and extend it to sum of arbitrary number of matrices case. Consider a linear weight update process defined as
\begin{equation}
\label{haha3}
{\boldsymbol{w}}\left( {i + 1} \right) = \left( {\sum\limits_{k = 1}^N {{\theta _k}\left( i \right){{\boldsymbol{B}}_k}} } \right){\boldsymbol{w}}\left( i \right)
\end{equation}
where ${\boldsymbol{w}}\in{\mathbb{R}}^{l\times 1}$. ${\theta}_1(i)$ is fixed at 1, while ${{\theta}}_k(i) \in {\mathbb{R}}, k=2,...,N$ are random variables with Gaussian distribution $N(0,1)$. By defining ${\boldsymbol{D}}\left( i \right) = E\left[ {{\boldsymbol{w}}\left( i \right){{\boldsymbol{w}}^T}\left( i \right)} \right]$ we can obtain following equation
\begin{equation}
\begin{aligned}
\!{\boldsymbol{D}}\left( {i + 1} \right) &=\! E\left[ {{\boldsymbol{w}}\left( {i + 1} \right){{\boldsymbol{w}}^T}\left( {i + 1} \right)} \right]\\
&=\! E\!\left[ {\left( {\sum\limits_{k = 1}^N\! {{\theta _k}\!\left( i \right){{\boldsymbol{B}}_k}} } \!\right){\boldsymbol{w}}\!\left( i \right){{\boldsymbol{w}}^T}\!\left( i \right)\left( {\sum\limits_{k = 1}^N\! {{\theta _k}\!\left( i \right){\boldsymbol{B}}_k^T} } \!\right)} \right] \\
&=\! \sum\limits_{k = 1}^N {{{\boldsymbol{B}}_k}{\boldsymbol{D}}\left( i \right){\boldsymbol{B}}_k^T}
\end{aligned}
\end{equation}
Further, using vectorization operator, we obtain
\begin{equation}
vec\left\{ {{\boldsymbol{D}}\left( {i + 1} \right)} \right\} = \sum\limits_{k = 1}^N {\left( {{{\boldsymbol{B}}_k} \otimes {{\boldsymbol{B}}_k}} \right)} vec\left\{ {{\boldsymbol{D}}\left( i \right)} \right\}
\end{equation}
it is known that if $\rho \left( {\sum\limits_{k = 1}^N {\left( {{{\boldsymbol{B}}_k} \otimes {{\boldsymbol{B}}_k}} \right)} } \right)< 1$, ${\boldsymbol{D}}\left( i \right)$ will converge to $0$ as $i\rightarrow\infty$ in both $l^2$ and $l$. Thus $E\left[ {{\boldsymbol{w}}\left( i \right)} \right]$ will also converge to $0$ as $i\rightarrow\infty$. Taking expectation of both sides of Eq.(\ref{haha3}), we can get
\begin{equation}
E\left[ {{\boldsymbol{w}}\left( {i + 1} \right)} \right] = {{\boldsymbol{B}}_1}E\left[ {{\boldsymbol{w}}\left( i \right)} \right]
\end{equation}
To ensure the convergence, $\rho\left({{\boldsymbol{B}}_1}\right)<1$ should be always guaranteed. Therefore, we have the following relation.
\begin{equation}
\label{haha1}
\rho \left( {\sum\limits_{k = 1}^N {\left( {{{\boldsymbol{B}}_k} \otimes {{\boldsymbol{B}}_k}} \right)} } \right)<1 \Rightarrow \rho \left( {{{\boldsymbol{B}}_1}} \right) < 1 \Rightarrow \rho \left( {{{\boldsymbol{B}}_1} \otimes {{\boldsymbol{B}}_1}} \right) < 1
\end{equation}
Moreover, assume that
\begin{equation}
\rho \left( {\sum\limits_{k = 1}^N {\left( {{{\boldsymbol{B}}_k} \otimes {{\boldsymbol{B}}_k}} \right)} } \right) = \varsigma
\end{equation}
where $\varsigma$ is arbitrary positive value. By defining $\beta  = \varsigma  + \varepsilon $ with arbitrary positive value $\varepsilon$, we will have
\begin{equation}
\rho \left( {\sum\limits_{k = 1}^N {\left( {\frac{{{{\boldsymbol{B}}_k}}}{{\sqrt \beta  }} \otimes \frac{{{{\boldsymbol{B}}_k}}}{{\sqrt \beta  }}} \right)} } \right) = \frac{\varsigma }{\beta } < 1
\end{equation}
Then, according to Eq.(\ref{haha1}),
\begin{equation}
\rho \left( {\frac{{{{\boldsymbol{B}}_1}}}{{\sqrt \beta  }} \otimes \frac{{{{\boldsymbol{B}}_1}}}{{\sqrt \beta  }}} \right) < 1
\end{equation}
which is equivalently
\begin{equation}
\label{haha2}
\rho \left( {{{\boldsymbol{B}}_1} \otimes {{\boldsymbol{B}}_1}} \right) < \varsigma  + \varepsilon
\end{equation}
Since Eq.(\ref{haha2}) always holds for arbitrary $\varepsilon$, we can obtain
\begin{equation}
\rho \left( {{{\boldsymbol{B}}_1} \otimes {{\boldsymbol{B}}_1}} \right) \le \rho \left( {\sum\limits_{k = 1}^N {\left( {{{\boldsymbol{B}}_k} \otimes {{\boldsymbol{B}}_k}} \right)} } \right)
\end{equation}
That's end the proof.

\subsection{Appendix C}
Rewrite ${\boldsymbol{\cal B}}$ as a block matrix
\begin{equation}
{\boldsymbol{\cal B}} = \left[ {\begin{array}{*{20}{c}}
{{{\boldsymbol{C}}_{11}}}&{{{\boldsymbol{C}}_{12}}}& \cdots &{{{\boldsymbol{C}}_{1l}}}\\
{{{\boldsymbol{C}}_{21}}}&{{{\boldsymbol{C}}_{22}}}& \cdots & \vdots \\
 \vdots & \vdots & \ddots &{{{\boldsymbol{C}}_{l - 1,l}}}\\
{{{\boldsymbol{C}}_{l1}}}& \cdots &{{{\boldsymbol{C}}_{l,l - 1}}}&{{{\boldsymbol{C}}_{ll}}}
\end{array}} \right]
\end{equation}
where ${\boldsymbol{C}}_{ij},i,j\in\{1,2,...,l\}$ are $l\times l$ matrices. Then ${{\boldsymbol{\cal B}}^{ \otimes {{\boldsymbol{I}}_N}}}$ can be derived as
\begin{equation}
{{\boldsymbol{\cal B}}^{ \otimes {{\boldsymbol{I}}_N}}} = \left[ {\begin{array}{*{20}{c}}
{{{\boldsymbol{I}}_N} \otimes {{\boldsymbol{C}}_{11}} \otimes {{\boldsymbol{I}}_N}}& \cdots &{{{\boldsymbol{I}}_N} \otimes {{\boldsymbol{C}}_{1l}} \otimes {{\boldsymbol{I}}_N}}\\
 \vdots & \ddots & \vdots \\
{{{\boldsymbol{I}}_N} \otimes {{\boldsymbol{C}}_{l1}} \otimes {{\boldsymbol{I}}_N}}& \cdots &{{{\boldsymbol{I}}_N} \otimes {{\boldsymbol{C}}_{ll}} \otimes {{\boldsymbol{I}}_N}}
\end{array}} \right]
\end{equation}
Assume $\xi$ is an eigenvalue of ${\boldsymbol{{\cal B}}}$, i.e.
${\boldsymbol{{\cal B} x}} = \xi {\boldsymbol{x}}$ where $\boldsymbol{x}$ is the corresponding eigenvector. Moreover, $\boldsymbol{x}$ can be rewrote as
\begin{equation}
{\boldsymbol{x}} = {\left[ {\begin{array}{*{20}{c}}
{{\boldsymbol{x}}_1^T}&{{\boldsymbol{x}}_2^T}& \cdots &{{\boldsymbol{x}}_l^T}
\end{array}} \right]^T}
\end{equation}
where ${{\boldsymbol{x}}_i}\in {\mathbb{R}}^{l\times1}$, $i=1,2,...,l$. We define a new vector ${{\boldsymbol{y}}_p}$
\begin{equation}
{{\boldsymbol{y}}_p} = vec\left\{ {\left[ {\begin{array}{*{20}{c}}
{{{\boldsymbol{x}}_1}}&{{{\boldsymbol{x}}_2}}& \cdots &{{{\boldsymbol{x}}_l}}
\end{array}} \right] \otimes {{\boldsymbol{Z}}_p}} \right\},p = 1,...,{N^2}
\end{equation}
where ${{\boldsymbol{Z}}_p}$ obeys that the $p$-th element of $vec\{{{\boldsymbol{Z}}_p}\}$ is 1 while other elements are set to 0. Therefore, one can get that following relation
\begin{equation}
{{\boldsymbol{B}}^{ \otimes {{\boldsymbol{I}}_N}}}{{\boldsymbol{y}}_p} = \xi {{\boldsymbol{y}}_p},p = 1,...,{N^2}
\end{equation}
That is, given arbitrary eigenvalue $\xi$ of ${\boldsymbol{\cal B}}$, ${{\boldsymbol{B}}^{ \otimes {{\boldsymbol{I}}_N}}}$ will have $N^2$ numbers of $\xi$ as the eigenvalues. Moreover, since the number of eigenvalues of ${\boldsymbol{\cal B}}$ is $l^2$ while ${{\boldsymbol{B}}^{ \otimes {{\boldsymbol{I}}_N}}}$ is $l^2\times N^2$, the eigenvalues of ${{\boldsymbol{B}}^{ \otimes {{\boldsymbol{I}}_N}}}$ will be
\begin{equation}
\{\underbrace {{\xi _1},{\xi _1},...,{\xi _1}}_{N^2},\underbrace {{\xi _2},{\xi _2},...,{\xi _2}}_{N^2},...,\underbrace {{\xi _{l^2}},{\xi _{l^2}},...,{\xi _{l^2}}}_{N^2}\}
\end{equation}
and consequently
\begin{equation}
\rho({\boldsymbol{\cal B}}^{\otimes \boldsymbol{I}_N})=\rho(\boldsymbol{\cal B})
\end{equation}
That's end the proof.

\subsection{Appendix D}
\par One can observe from of Eq.(\ref{Fs}) that
\begin{equation}
\begin{aligned}
{\boldsymbol{V}}&={\left(1 - \frac{\mu}{M} \right)^2 \boldsymbol{I}}+ {\frac{{N + 1}}{{{M^2}}}\mu^2\sum\limits_{k = 1}^M {\left( {{{\boldsymbol{ T}}_k} \otimes {{\boldsymbol{ T}}_k}} \right)} }\\
&={\left(1 - \frac{\mu}{M} \right)^2 \boldsymbol{I}}+ {\frac{{N + 1}}{{{M^2}}}\mu^2 diag(vec\{\boldsymbol{S}^T\boldsymbol{S}\})}
\end{aligned}
\end{equation}
is an diagonal matrix and ${{\boldsymbol{ A}}_2^{\otimes}}=\left( {{{\boldsymbol{ A}}_2} \otimes {{\boldsymbol{ A}}_2}} \right)$ is a column stochastic matrix where the sum of each column is equal to 1. Thus,
according to the theorem that the spectral radius is not more than any norms of the matrix, we will have
\begin{equation}
\begin{aligned}
\rho({\boldsymbol{F}})&=\rho({\boldsymbol{V}}{{\boldsymbol{ A}}_2^{\otimes}})\\
&=\rho({{\boldsymbol{ A}}_2^{\otimes}}{\boldsymbol{V}})\\
&\leq\max\{diag({\boldsymbol{V}})\}\\
&={\left(1 - \frac{\mu}{M} \right)^2 \boldsymbol{I}}+ \frac{{N + 1}}{{{M^2}}}\mu^2\max\{{{\boldsymbol{S}}^T}{{\boldsymbol{S}}}\}
\end{aligned}
\end{equation}
Therefore, if $\max\{diag({\boldsymbol{V}})\}<1$, then $\rho({\boldsymbol{F}})<1$ will be guaranteed. Thus the sufficient condition for $\rho({\boldsymbol{F}})<1$ will be
\begin{equation}
\label{musufficient}
{\mu}<\frac{2M}{(N+1)\max\{{{\boldsymbol{S}}^T}{{\boldsymbol{S}}}\}+1}
\end{equation}
To proof the right inequality of Eq.(\ref{mumax}), we first propose the following theorem
\begin{myTheo}
For arbitrary real square matrices sequence $\{\boldsymbol{B}_k\}_{k=1}^{t}\in\mathbb{R}^{l\times l},t,l\in\boldsymbol{N}_+$, the following inequality will always hold
\begin{equation}
\begin{aligned}
\rho &\left( {p\sum\limits_{k = 1}^N {{{\boldsymbol{B}}_k}}  \otimes \sum\limits_{k = 1}^N {{{\boldsymbol{B}}_k}}  + q\sum\limits_{k = 1}^N {\left( {{{\boldsymbol{B}}_k} \otimes {{\boldsymbol{B}}_k}} \right)} } \right) \\
&\ge \left( {p + \frac{q}{N}} \right)\rho \left( {\sum\limits_{k = 1}^N {{{\boldsymbol{B}}_k}}  \otimes \sum\limits_{k = 1}^N {{{\boldsymbol{B}}_k}} } \right)
\end{aligned}
\end{equation}
where the equality will always hold when ${{{\boldsymbol{B}}_1}}={{{\boldsymbol{B}}_2}}=...={{{\boldsymbol{B}}_N}}$. The proof is given in Appendix D.
\end{myTheo}
Since $\sum_{k=1}^{P}{\boldsymbol{T}}_k={\boldsymbol{I}}$, according to Theorem 3 we can obtain
\begin{equation}
\rho({\boldsymbol{F}})\geq\left(1-2\frac{\mu}{M}+\frac{\mu^2}{M^2}+\frac{(N+1)\mu^2}{M^2P}\right)\rho\left( {{\boldsymbol{A}}_2}\!\otimes\!{{\boldsymbol{A}}_2} \right)
\end{equation}
It is known that $\rho\left({\boldsymbol{A}}_2\right)=1$ since ${\boldsymbol{A}}_2$ are both column stochastic matrices. Therefore $\rho\left( {{\boldsymbol{A}}_2}\!\otimes\!{{\boldsymbol{A}}_2} \right)=1$ and
\begin{equation}
\rho({\boldsymbol{F}})\geq1-2\frac{\mu}{M}+\frac{\mu^2}{M^2}+\frac{(N+1)\mu^2}{M^2P}
\end{equation}
The equality will hold if $\boldsymbol{T}_1=\boldsymbol{T}_2=...=\boldsymbol{T}_P$. If $1-2\frac{\mu}{M}+\frac{\mu^2}{M^2}+\frac{(N+1)\mu^2}{M^2P}\geq1$, $\rho({\boldsymbol{F}})$ will always no less than 1. Thus one can obtain the necessary condition for the convergence
\begin{equation}
\label{munecessary}
\mu<\frac{2PM}{P+N+1}
\end{equation}
Combining the sufficient condition in Eq.(\ref{musufficient}) and necessary condition in Eq.(\ref{munecessary}), we can obtain the range of upper bound $\mu_{max}$
\begin{equation}
\frac{2M}{(N+1)\max\{{{\boldsymbol{S}}^T}{{\boldsymbol{S}}}\}+1}\leq{\mu_{max}}\leq\frac{2PM}{P+N+1}
\end{equation}
That's end the proof.


\subsection{Appendix D}
For arbitrary real square matrices sequence $\{\boldsymbol{B}_k\}_{k=1}^{t}\in{\mathbb{R}}^{l\times l},t,l\in\boldsymbol{N}_+$, we will have
\begin{equation}
\begin{aligned}
&p\sum\limits_{k = 1}^N {{{\boldsymbol{B}}_k}}  \otimes \sum\limits_{k = 1}^N {{{\boldsymbol{B}}_k}}  + q\sum\limits_{k = 1}^N {\left( {{{\boldsymbol{B}}_k} \otimes {{\boldsymbol{B}}_k}} \right)} \\
& = p\sum\limits_{k = 1}^N {{{\boldsymbol{B}}_k}}  \otimes \sum\limits_{k = 1}^N {{{\boldsymbol{B}}_k}} \\
&~~~ +\! \frac{q}{N}\!\left( {\sum\limits_{k = 1}^N \!{{{\boldsymbol{B}}_k}}  \!\otimes\! \sum\limits_{k = 1}^N\! {{{\boldsymbol{B}}_k}}  \!+\! \sum\limits_{i = 1}^N \!{\sum\limits_{j = i + 1}^N\! \!\!{\left( {{{\boldsymbol{B}}_i} \!-\! {{\boldsymbol{B}}_j}} \right) \!\otimes\! \left( {{{\boldsymbol{B}}_i} \!-\! {{\boldsymbol{B}}_j}} \right)} } } \!\!\right)\\
& = \left( {p + \frac{q}{N}} \right)\sum\limits_{k = 1}^N {{{\boldsymbol{B}}_k}}  \otimes \sum\limits_{k = 1}^N {{{\boldsymbol{B}}_k}}  \\
&~~~+ \frac{q}{N}\sum\limits_{i = 1}^N {\sum\limits_{j = i + 1}^N {\left( {{{\boldsymbol{B}}_i} - {{\boldsymbol{B}}_j}} \right) \otimes \left( {{{\boldsymbol{B}}_i} - {{\boldsymbol{B}}_j}} \right)} }
\end{aligned}
\end{equation}
According to Theorem 1, we can obtain
\begin{equation}
\begin{aligned}
\rho &\left( {p\sum\limits_{k = 1}^N {{{\boldsymbol{B}}_k}}  \otimes \sum\limits_{k = 1}^N {{{\boldsymbol{B}}_k}}  + q\sum\limits_{k = 1}^N {\left( {{{\boldsymbol{B}}_k} \otimes {{\boldsymbol{B}}_k}} \right)} } \right) \\
&\ge \left( {p + \frac{q}{N}} \right)\rho \left( {\sum\limits_{k = 1}^N {{{\boldsymbol{B}}_k}}  \otimes \sum\limits_{k = 1}^N {{{\boldsymbol{B}}_k}} } \right)
\end{aligned}
\end{equation}
Specifically, when ${{{\boldsymbol{B}}_1}}={{{\boldsymbol{B}}_2}}=...={{{\boldsymbol{B}}_N}}$ the equality will hold.
That's end the proof.

\end{appendices}

\bibliographystyle{IEEEtran}
\bibliography{DLMS_CS_IEEE}

\end{document}